\documentclass{emulateapj}
\usepackage{amsmath}    
\usepackage{color}

\def\Dwa{$\,$\uppercase\expandafter{\romannumeral5}$\,$}

\def\sles{\lower2pt\hbox{$\buildrel {\scriptstyle <}
   \over {\scriptstyle\sim}$}}

\def\sgreat{\lower2pt\hbox{$\buildrel {\scriptstyle >}
   \over {\scriptstyle\sim}$}}
\def\sharpnull#1{}

\newcommand{\iso}[2]{\ensuremath{^{#1}\rm{#2}}}

\newcommand{\nifs}{\iso{56}{Ni}}
\newcommand{\cofs}{\iso{56}{Co}}
\newcommand{\fefs}{\iso{56}{Fe}}

\newcommand{\tsim}{\ensuremath{\sim}}
\newcommand{\ra}{\ensuremath{\rightarrow}}

\newcommand{\msun}{\ensuremath{M_{\odot}}}

\null\voffset=-1.0pc  

\begin{document}
\def \jcp {J. of Comp. Phys.}
\def \m {$ M_\odot$}
\def \l {$ L_\odot$}
\def \ro {g/cm$^{3}$}
\def \f {{\nu}}
\def \pos {{\mathbf r}}
\def \vel {{\mathbf v}}
\def \dir {{\mathbf \Omega}}
\def \dphidt {{\partial \Phi \over {\partial t}}}
\def \bn {{\mathbf n}}
\def \bB {{\mathbf B}}
\def \bBovr {(\frac{\mathbf B}{\rho})}
\def \bv {{\mathbf v}}
\def \Bf {{B_{\phi}}}
\def \vf {{v_{\phi}}}
\def \half {\frac{1}{2}}
\def \ovr  {\frac{1}{r}}
\def \dpdr {{\partial p \over {\partial r}}}
\def \dpdz {{\partial p \over {\partial z}}}
\def \Ni56 {$^{56}$Ni }
\def \Ti44 {$^{44}$Ti }
\def \Ca40 {$^{40}$Ca }
\def \Cr48 {$^{48}$Cr }

\newcommand{\der}[2]{\frac{D#1}{D#2}}
\newcommand{\Dert}[1]{\frac{d#1}{dt}}
\newcommand{\dpar}[2]{\frac{\partial#1}{\partial#2}}
\newcommand{\dpart}[1]{\frac{\partial#1}{\partial t}}
\newcommand{\E}[2]{$#1\times 10^{#2}$}

\slugcomment{\bf}
\slugcomment{Submitted to Ap.J.}

\title{Helium Shell Detonations on Low Mass White Dwarfs as a Possible Explanation for SN~2005E}

\author{Roni Waldman\altaffilmark{1}, Daniel Sauer\altaffilmark{2}, Eli Livne\altaffilmark{1}, Hagai Perets\altaffilmark{3}, Ami Glasner \altaffilmark{1},         Paolo Mazzali \altaffilmark{4} \altaffilmark{5} \altaffilmark{6}, James W. Truran \altaffilmark{7} \altaffilmark{8}, Avishay Gal-Yam \altaffilmark{9}}

\altaffiltext{1}{Racah Institute of Physics, The Hebrew University, Jerusalem 91904,  Israel}
\altaffiltext{2}{Stockholm University, Department for Astronomy, AlbaNova University Center, 106 91 Stockholm, Sweden; Current Address: Meteorological Institute, Ludwig-Maximilians-University, Theresienstr. 37, 80333 Munich, Germany}
\altaffiltext{3}{Harvard-Smithsonian Center for Astrophysics, 60 Garden St., Cambridge MA, USA 02138}
\altaffiltext{4}{Max-Planck Institut f\"ur Astrophysik, Karl-Schwarzschild-Str. 1, 85748 Garching, Germany}
\altaffiltext{5}{Scuola Normale Superiore, Piazza dei Cavalieri 7, 56126 Pisa, Italy}
\altaffiltext{6}{Istituto Naz. di Astrofisica-Oss. Astron., vicolo dell'Osservatorio, 5, 35122 Padova, Italy}
\altaffiltext{7}{Physics Division, Argonne National Laboratory, Argonne, IL 60439}
\altaffiltext{8}{Department of Astronomy and Astrophysics, Enrico Fermi Institute,
   and Joint Institute for Nuclear Astrophysics, University of
   Chicago, 5640 South Ellis Avenue, Chicago, IL 60637}
\altaffiltext{9}{Benoziyo Center for Astrophysics, Faculty of Physics, The Weizmann Institute of Science, Rehovot 76100, Israel}

\begin{abstract}

Recently several type Ib supernovae (SNe; with the prototypical SN~2005E) have been shown to have atypical properties.
These SNe are faint (absolute peak magnitude of $\sim-15$) and fast SNe that show unique composition. They are inferred to have low ejecta mass (a few tenths of a solar mass) and to be highly enriched in calcium, but poor in silicon elements and nickel. These SNe were therefore suggested to belong to a new class of calcium-rich faint SNe explosions. Their properties were proposed to be the result of helium detonations that may occur on helium accreting white dwarfs.
In this paper we theoretically study the scenario of helium detonations, and focus on the results of detonations in accreted helium layers on low mass carbon-oxygen (CO) cores. We present new results from one dimensional simulations of such explosions, including their light curves and spectra. We find that when the density of the helium layer is low enough the helium detonation produces large amounts of intermediate elements, such as calcium and titanium, together with a large amount of unburnt helium. Our results suggest that the properties of calcium-rich faint SNe could indeed be consistent with the helium-detonation scenario on small CO cores. Above a certain density (larger CO cores) the detonation leaves mainly \iso{56}Ni and unburnt helium, and the predicted spectrum will unlikely fit the unique features of this class of SNe.
Finally, none of our studied models reproduces the bright, fast evolving light curves of another type of peculiar SNe suggested to originate in helium detonations (SNe 1885A, 1939B and 2002bj).
\end{abstract}

\keywords{nucleosynthesis, hydrodynamics, supernovae}

\section{Introduction}
\label{intro}

Recently, a new type of peculiar type Ib supernova (SN) has been discovered (with the prototypical SN~2005E and a full sample of 8 SNe; \citealp{perets10}). These helium rich SNe show several intriguing features. They show fast and faint light curves (B-band peak luminosity of -15) and dominant lines of calcium in the nebular spectrum. Only a small fraction of radioactive nickel is found in their ejecta, and Si group elements seem to be completely absent in their nebular spectra. In addition, the environment of these SNe is typically old \citep{perets10,perets10b}; inconsistent with massive young progenitors, usually thought to be associated with type Ib SNe.
It was therefore proposed that these objects may result from helium detonations that occur on helium accreting white dwarfs.
 Energetically, the moderate observed velocities of $11,000~{\rm km~sec^{-1}}$ together with the estimated ejected mass (a few tenths of a solar mass), are consistent with the binding energy of $0.2\,\msun$ of helium.

Other peculiar low mass and even faster evolving SNe, but bright (peak luminosity of $\sim-18.5$) have also been studied \citep{dev85,chevalier88,poznanski10,perets10c}. These SNe have also been suggested to result from helium detonations. Although we shall remark on these SNe, their light curves are brighter and faster than those produced in our simulations, and are unlikely to be produced by these scenarios.

In the past decades, several authors discussed the scenario of detonations in helium layers, accreted on carbon-oxygen cores, suggesting ``peculiar'' SNe with faint, fast light curves \citep[see e.g.][]{woosley80,nomoto80,nomoto82,woosley86,L1,L2,L3,W2,L4}. There are several main issues here which deserve close investigation.
The first issue is the explosion mechanism and especially the question whether a detonation which
develops first in the helium layer could ignite a second (successive) detonation in the CO core. The second important issue includes the nucleosynthesis and its impact on the shape of both the light curve and spectrum.  The third issue is the evolution of helium accreting white dwarfs to thermonuclear runaway.

The question whether a second detonation occurs has been discussed many times. If this happens, the process will lead to the disruption of the entire star, and the consequences
will be completely different from those obtained in the case where the core is not burning. Livne \& Glasner \citep{L1,L2,L3} have studied this, where a small reaction network of 13 species has been used. Under the assumption of spherical symmetry,
the CO core always experiences ignition at the center, after a converging pressure wave, which propagates
from the core-helium boundary, reaches the center \citep{L1,W2}. However, in a more realistic
scenario, helium detonation is unlikely to start at a spherical shell simultaneously. Rather it is more likely
to ignite at a small area near the interface, which can be approximated by a point ignition. The problem
becomes a two dimensional problem with cylindrical symmetry around an axis defined by the line connecting
the center of the star and the ignition point. \cite{L2,L3} have shown
two ways by which the core can be ignited in this case. Depending on actual parameters (mainly the density
of the fuels near the interface), the sliding helium detonation emits a strong oblique shock into the adjacent
CO core, which in some cases is strong enough to drive a CO detonation near the interface. Otherwise,
converging shock waves propagate inwards and eventually converge on the axis, but in this case off center.
Consequently, driving a second CO detonation from that convergence point is also probable if the density there is not
too low. Note however that those results were obtained for rather massive CO cores of $0.85\,\msun$ and above.

 The main study of the second important issue, namely the role of nucleosynthesis in the explosion and its appearance, has been done by \cite{W2},
who performed a series of one dimensional simulations using a detailed reaction network. They
used CO cores between $0.6\,\msun$ and $0.9\,\msun$ with helium mantles between $0.13\,\msun$ and $0.63\,\msun$ (where the total was kept
below the Chandrasekhar limit). In agreement with \cite{L1}, they find that the core is being ignited
in all cases by converging pressure waves. A common important observable feature of those models is their
very fast light curve, which rises to maximum over roughly 10-12 days. The amounts of \iso{56}Ni vary among
models, between $0.2\,\msun$ and $0.98\,\msun$, with strong correlation between the core mass and the amount of \iso{56}Ni produced.
Moreover, significant amounts of calcium and titanium isotopes are being synthesized. Interestingly, similar results
were obtained by 2D simulations \citep{L4}, where similar models were simulated under the assumption of point ignition.
A more comprehensive study in both 2D and 3D was recently carried out by \cite{fink07}. Their results confirm the main conclusions of the previous multi-dimensional studies.

Finally, the evolution of helium accreting white dwarfs to thermonuclear runaway is a very complicated subject.
Moreover, the onset of helium detonation at the base of such helium layers is a speculative process, yet to be
explored. Some progress has been recently reported by \cite{shen09}. In the context of our study, their main
result is the minimal helium mass required for dynamical burning as function of core mass (Fig. 5 there).
They also point out that the exact composition of the accreted helium plays an important role in the runaway. Previous results, published by \cite{Y1} are consistent with the above.
However, they focused on the possible effects of rotation and found that rotation may inhibit the runaway.

In this paper we ignore most of the above complications and repeat the work of \cite{W2},
but extend the range of core masses to lower values. Contrary to previous work, which focused on the higher edge of the
mass range, we are interested here in low mass progenitors which presumably can reproduce the observations of SN~2005E-like objects. Note that part of the phase space of low mass cores was also recently (and independently) studied by \cite{shen10}, as we shall briefly discuss later.

\section{Explosion models}
\subsection{Tools and initial configurations}
\label{tools}

We use the hydro code Vulcan/1D (V1D) which consists of an explicit Lagrangian scheme.
The reaction network is based upon REACLIB with 160 elements between hydrogen and Ni, including neutrons, \iso{1,2}H, \iso{3,4}He, \iso{7}Li, \iso{7}Be, \iso{8}B, \iso{12-14}C, \iso{13-15}N, \iso{14-18}O, \iso{17-19}F, \iso{18-22}Ne, \iso{20-23}Na, \iso{21-26}Mg, \iso{23-27}Al, \iso{24-30}Si, \iso{27-31}P, \iso{29-34}S, \iso{31-37}Cl, \iso{33-38}Ar, \iso{36-39}K, \iso{37-44}Ca, \iso{40-45}Sc, \iso{41-50}Ti, \iso{44-51}V, \iso{45-54}Cr, \iso{48-55}Mn, \iso{49-58}Fe, \iso{50-59}Co, \iso{53-64}Ni \citep{W1}.
The initial models are hydrostatic configurations, consisting of CO cores, with masses of 0.45, 0.5 and $0.6\,\msun$, and He layers of 0.15, 0.2 and $0.3\,\msun$. The CO core consists of pure, equal mass fractions of \iso{12}C and \iso{16}O, while the He layer is pure \iso{4}He (except for the model discussed in \ref{nucleosynthesis}, which had \iso{12}C mixed into the He layer). The CO core is assumed to be isentropic,
with central temperature of $10^7 {\rm K}$ and density that fits the given core mass. The helium
layer is also isentropic, with bottom temperature of $2\times10^8 {\rm K}$ \citep{Y1,shen09}.
Our temperatures are a bit higher than those used in \cite{W2} and therefore our densities are slightly
lower. The models are well resolved with more than 1000 spatial zones, where the CO core is represented by $\approx 500$ zones, gradually growing from $0.001\,\msun$ at the center to $0.005\,\msun$ at $0.05\,\msun$ below the edge of the core, then gradually decreasing to $10^{-4}\,\msun$ and remaining at this value throughout the He layer. In Table~\ref{tab:models} we detail
the hydrostatic parameters of the cases we have simulated.

The detonation at the base of the helium layer is driven artificially by giving
20 zones large positive radial velocity of $10^9\,{\rm cm\,s^{-1}}$. Usually this drives a detonation immediately,
but, at very low density the detonation may die out after a short period of time. This
suggests that the question of how/when helium detonation may be ignited spontaneously,
should be studied separately under appropriate conditions and smaller scales.
In any case, when formed, the self sustained detonation is rather weak, leading to incomplete
helium burning over most of the layer. As mentioned earlier, spherically symmetric simulations
lead always to CO ignition at the center. To avoid this complication we artificially inhibit
here the burning in the core. This is temporarily justified by the fact that in the 2D case the
convergence of those waves is off center, and may not ignite the core when the density at the
convergence region is low enough [see \cite{L2,L3} for details].

The simulation is run without rezoning until the shock wave approaches the center (typically at $\simeq 0.1\,\msun$ from the center, $\simeq 1\,{\rm s}$ after the detonation is ignited, see Fig.~\ref{fig:profiles}). From then on, as the He layer draws away from the core, the core is rezoned more and more coarsely. This is done in order to prevent the shocks, that are going back and forth through the core, from unnecessarily diminishing the time step (this does not affect the dynamics of the He layer).
 The simulation is run until $10^5\,{\rm s}$, at which epoch the densities in the He layer are low enough so that radiative transfer calculations can be carried out. Nuclear reactions are followed only above a temperature of $10^7\,{\rm K}$, whereas weak interaction decays are followed throughout the simulation for all zones in the helium layer.

\subsection{Results}

\begin{figure*}
\begin{minipage}[t]{0.5\linewidth}
\centering
\includegraphics[width=1.\linewidth]{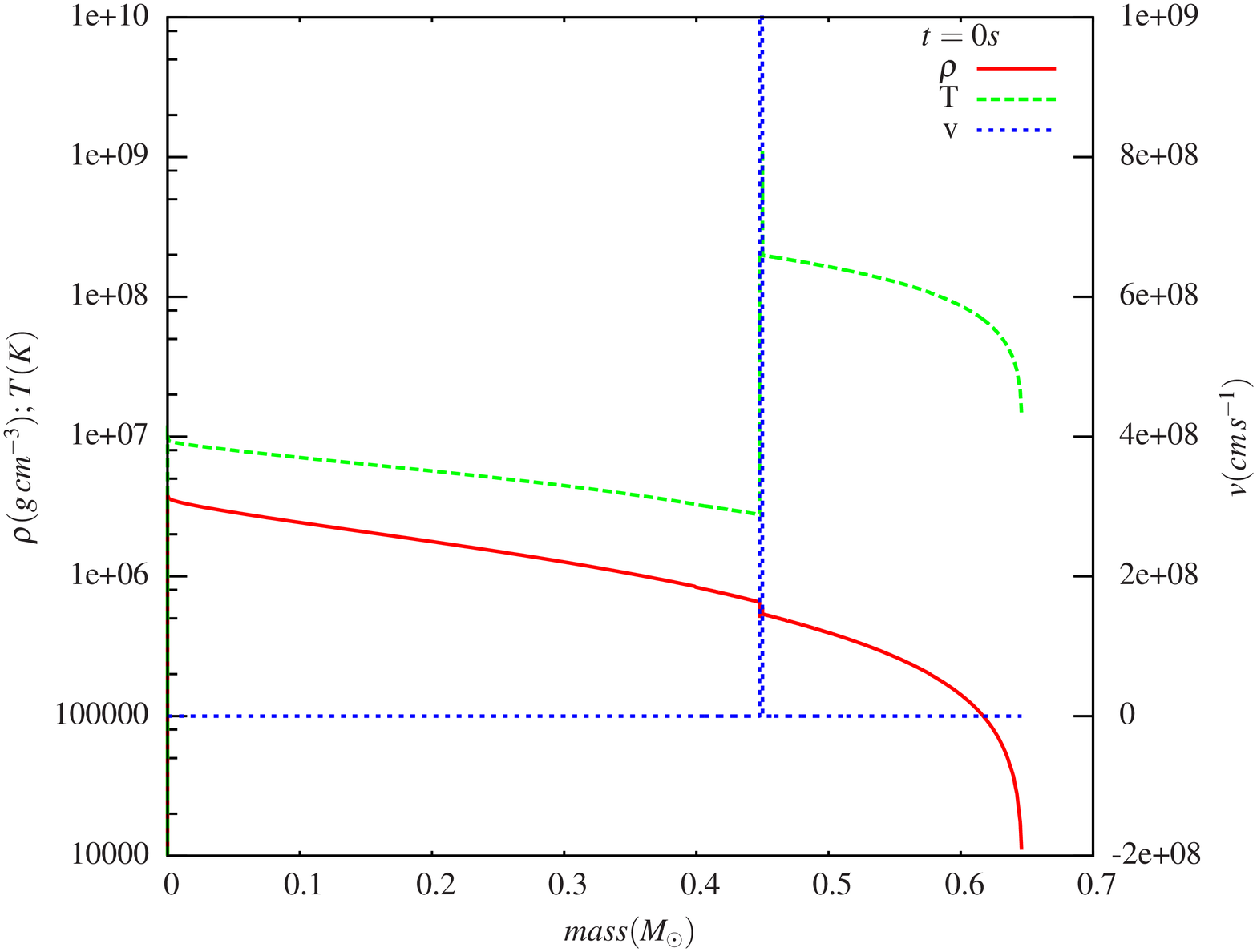}
\end{minipage}
\begin{minipage}[t]{0.5\linewidth}
\centering
\includegraphics[width=1.\linewidth]{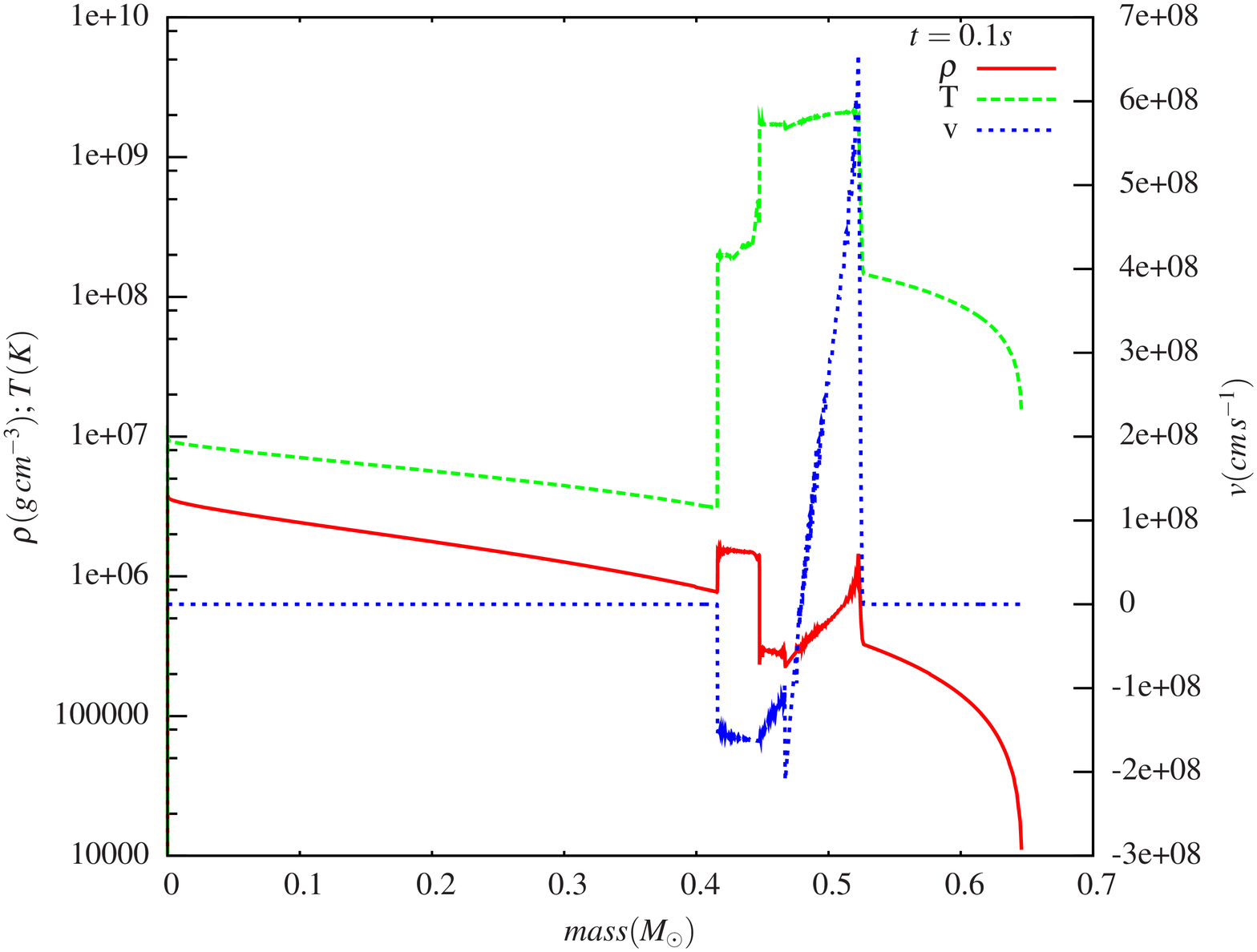}
\end{minipage}
\begin{minipage}[t]{0.5\linewidth}
\centering
\includegraphics[width=1.\linewidth]{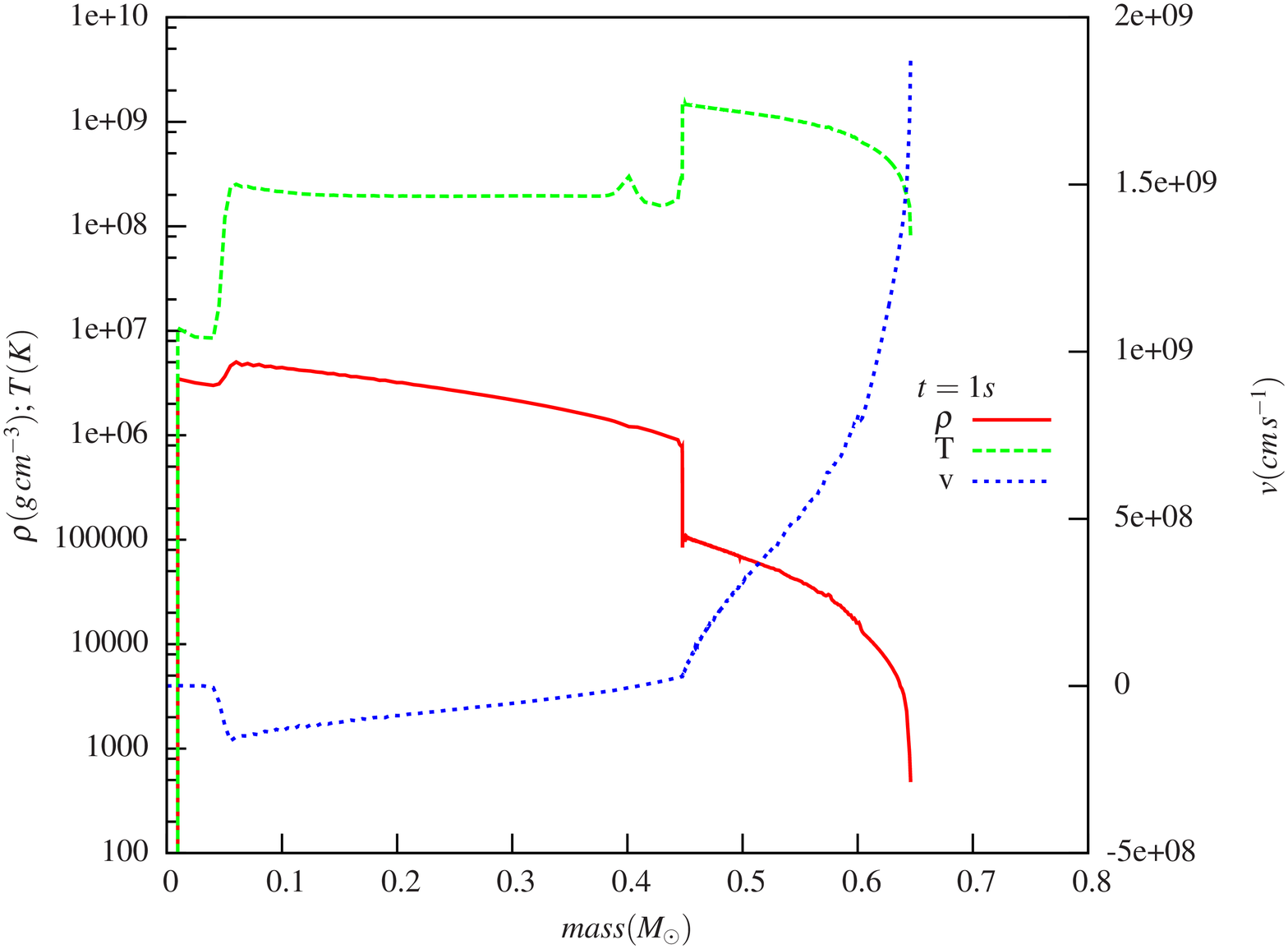}
\end{minipage}
\begin{minipage}[t]{0.5\linewidth}
\centering
\includegraphics[width=1.\linewidth]{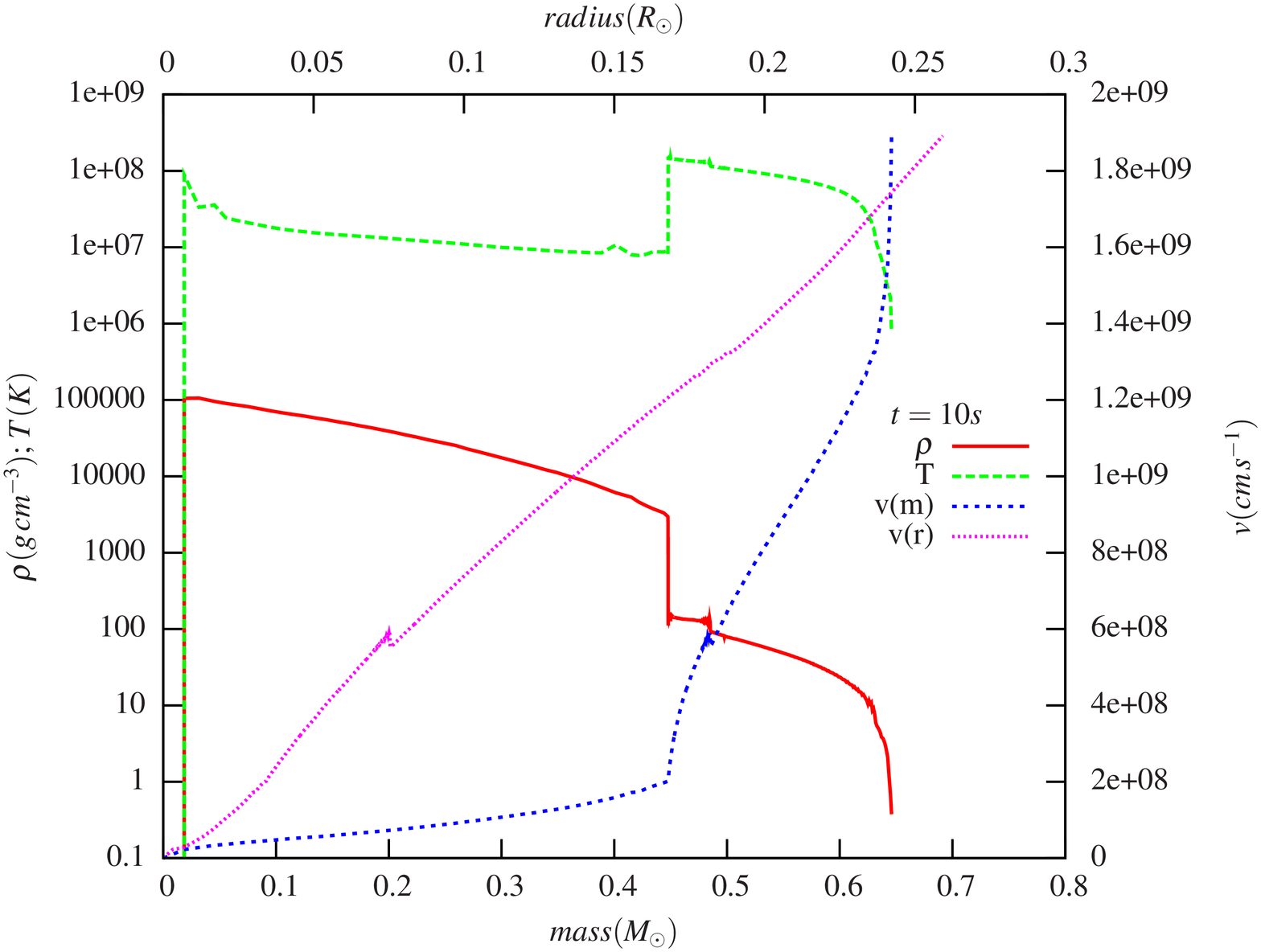}
\end{minipage}
\begin{minipage}[t]{0.5\linewidth}
\centering
\includegraphics[width=1.\linewidth]{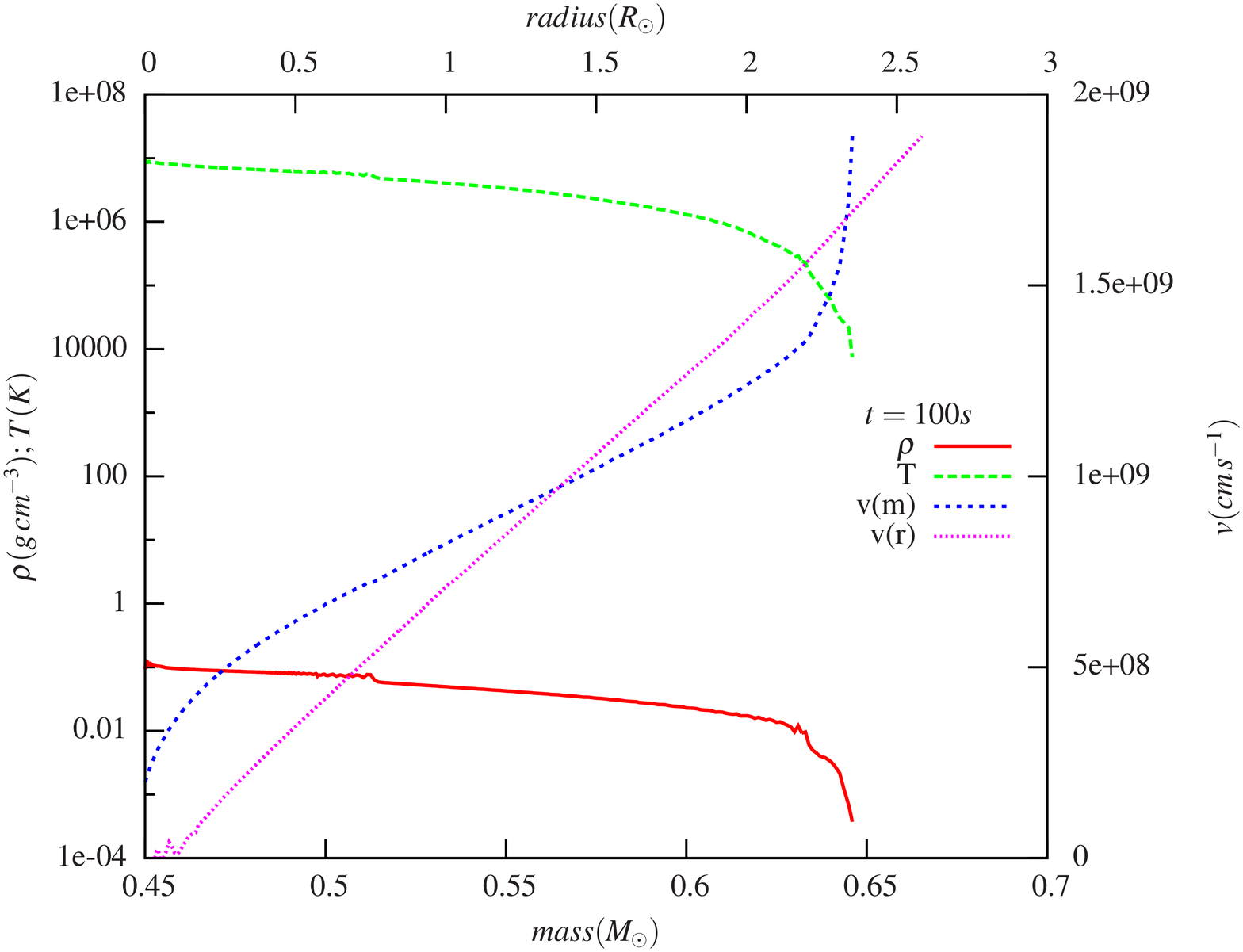}
\end{minipage}
\begin{minipage}[t]{0.5\linewidth}
\centering
\includegraphics[width=1.\linewidth]{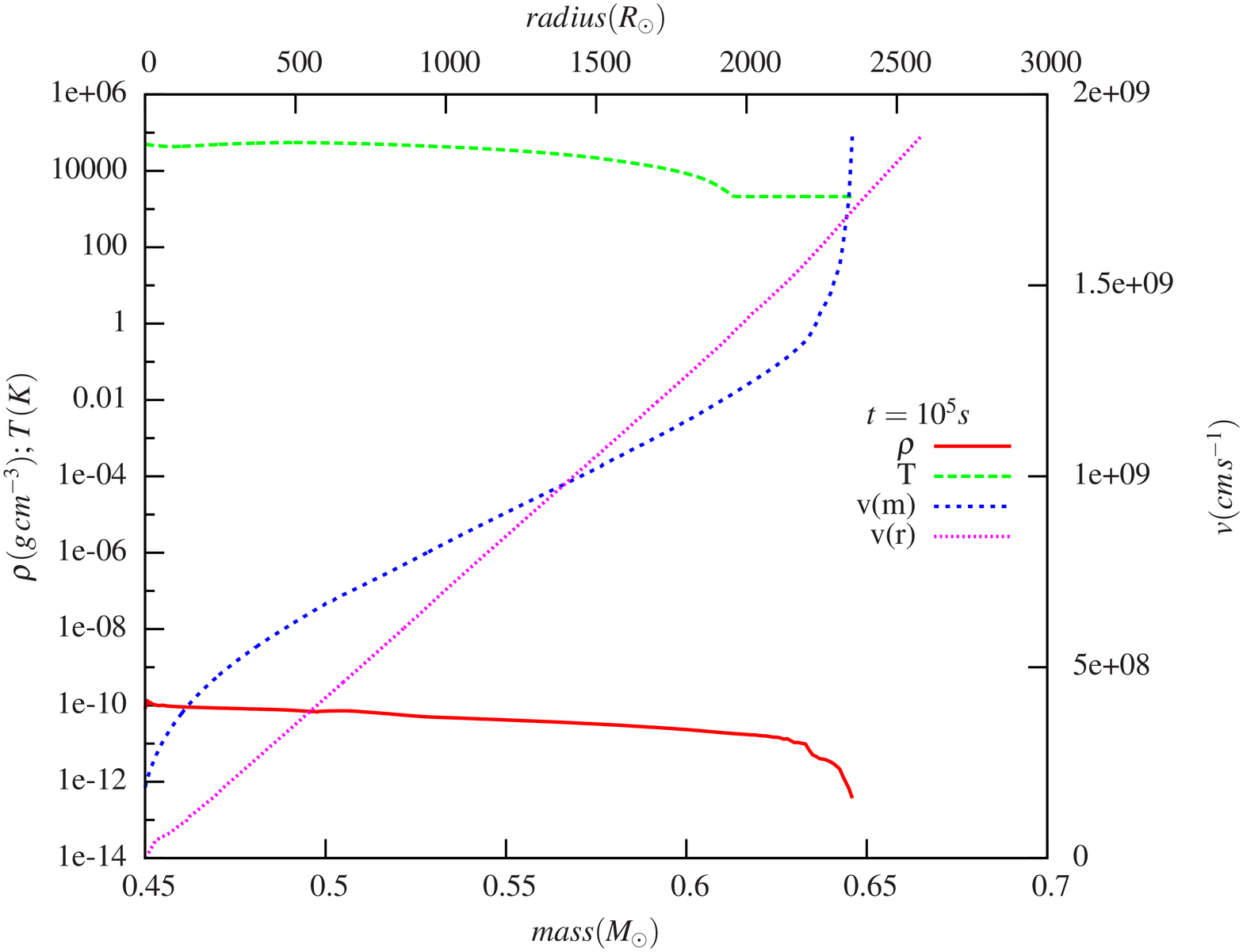}
\end{minipage}
\caption{Evolution of density (solid red line), temperature (long-dashed green line) and velocity (short-dashed blue line) profile of model CO.45He.2, having a CO core of $0.45\,\msun$ and He layer of $0.2\,\msun$. From $t=10\,{\rm s}$, the velocity is homologous, as shown against the radius (dotted purple line). }
  \label{fig:profiles}
\end{figure*}

The evolution of the density, temperature, and velocity profiles of a typical model CO.45He.2, having a CO core of $0.45\,\msun$ and He layer of $0.2\,\msun$, is shown in Fig.~\ref{fig:profiles}. The initial velocity of $10^9\,{\rm cm\,s^{-1}}$  at $t=0$ injected at the base of the He layer (in order to initialize the detonation) can be seen. At $t=0.1\,{\rm s}$, a detonation front has already formed, the maximum temperature in the He layer is $1.8 \times 10^9\,{\rm K}$. At $t=1\,{\rm s}$, the outward going shock has swept through all of the WD, while the inward going shock is approaching the center. The maximum temperature in the He layer at this stage is $1.5\times 10^9\,{\rm K}$, whereas in the core it reaches $2.5 \times 10^8\,{\rm K}$. It is worthwhile to mention that when the shock reaches the center, the temperature there sharply rises to $\approx 1.3 \times 10^9\,{\rm K}$. At $t=10\,{\rm s}$, the velocity in the He layer is nearly homologous, and nuclear burning has almost ceased, as the maximum temperature in the He layer has dropped to $1.5\times 10^8\,{\rm K}$. From then on, as can be seen for $t=100\,{\rm s}$ and $t=10^5\,{\rm s}$, the He layer continues expanding homologously, nuclear reactions being exclusively weak-interaction decays.

\begin{figure*}
\begin{minipage}[t]{0.5\linewidth}
\centering
\includegraphics[width=1.\linewidth]{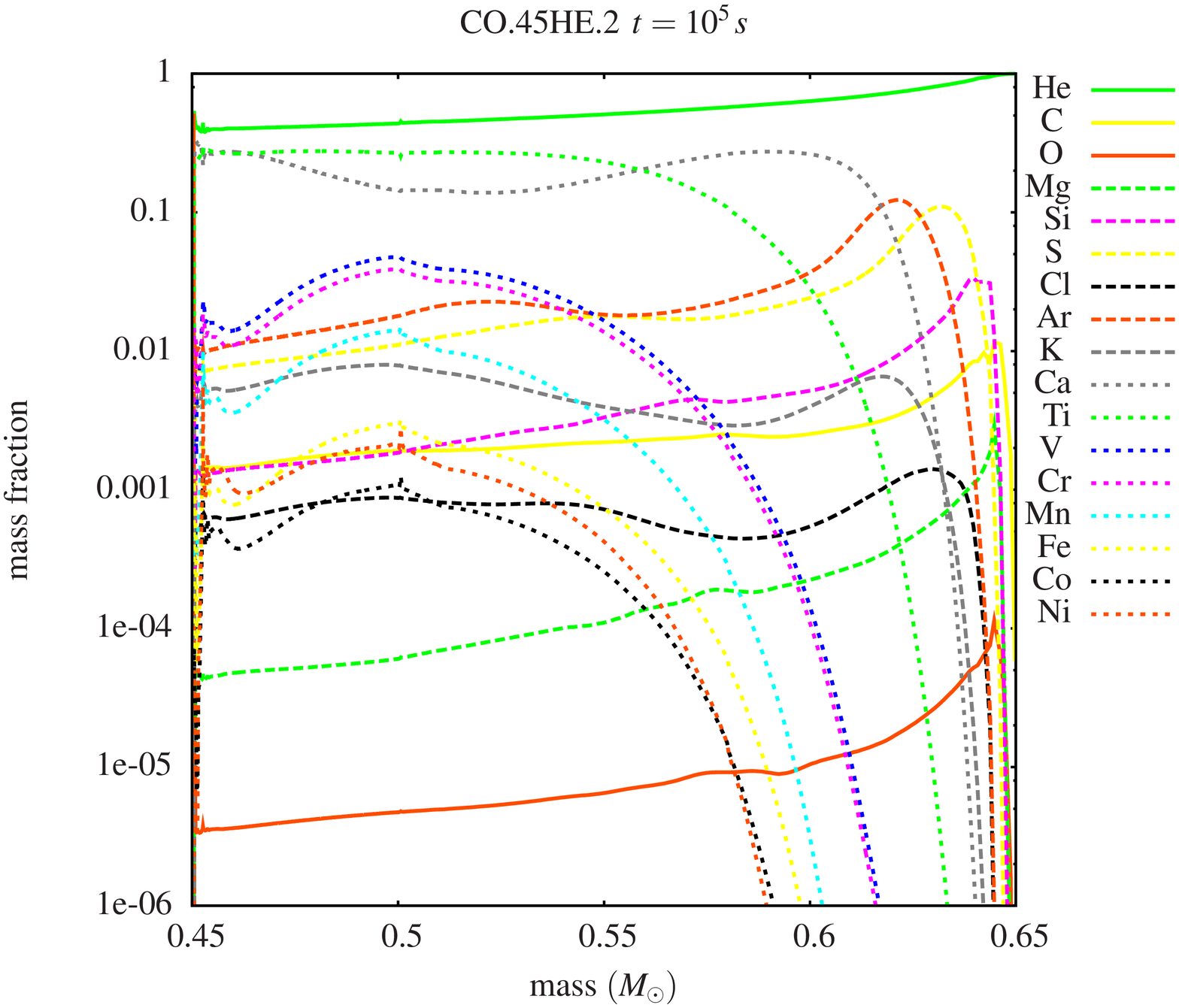}
\end{minipage}
\begin{minipage}[t]{0.5\linewidth}
\centering
\includegraphics[width=1.\linewidth]{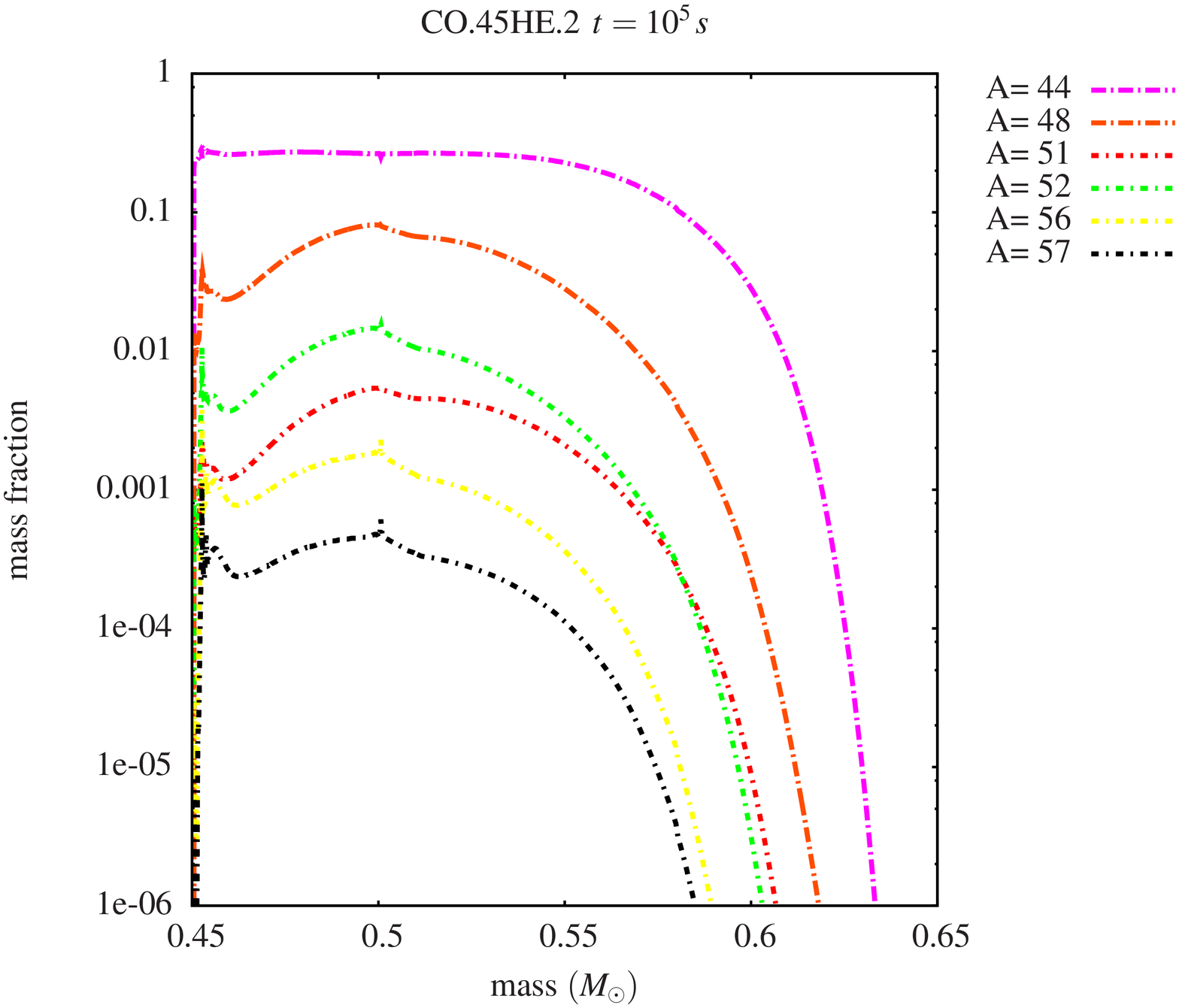}
\end{minipage}
\begin{minipage}[t]{0.5\linewidth}
\centering
\includegraphics[width=1.\linewidth]{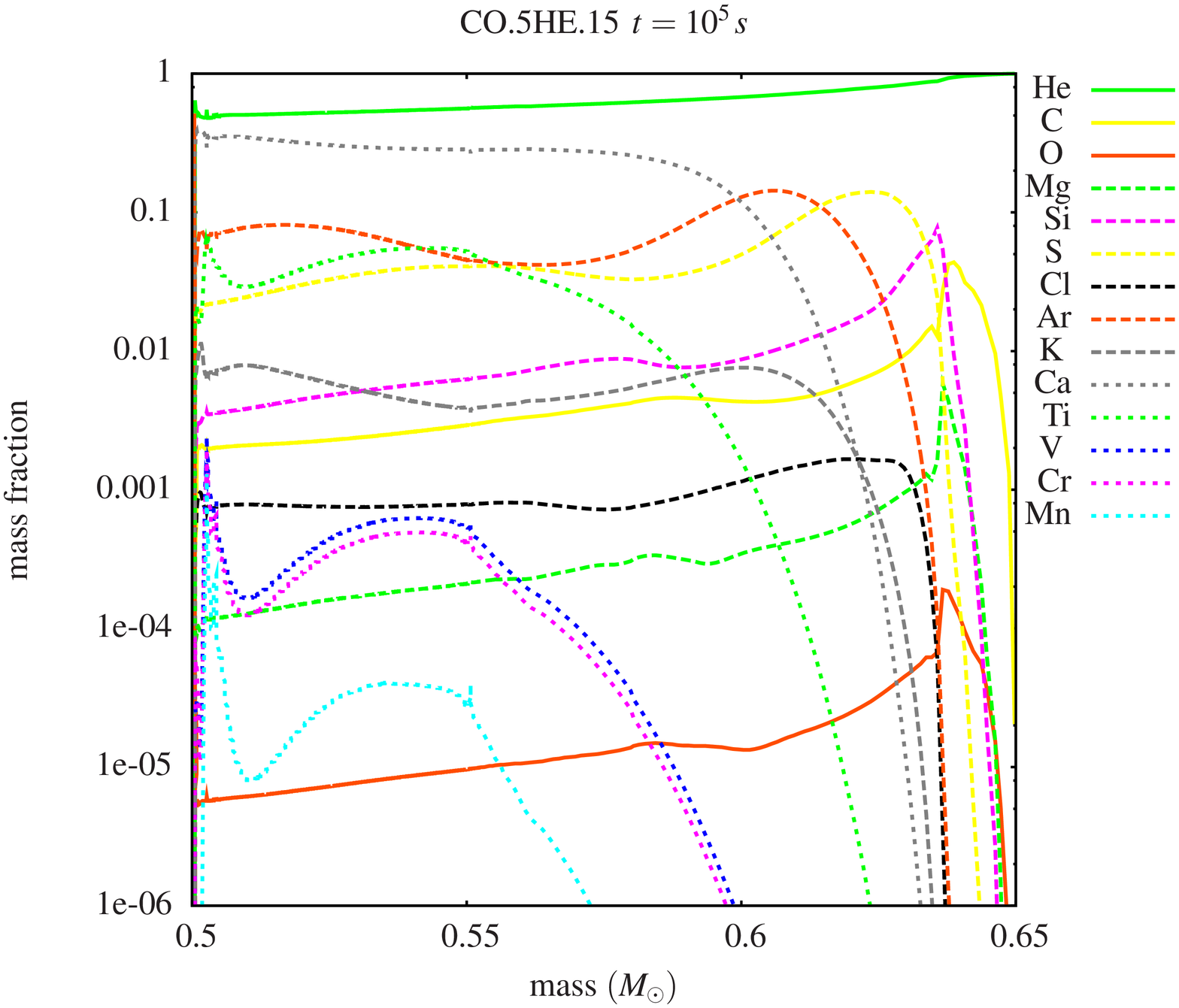}
\end{minipage}
\begin{minipage}[t]{0.5\linewidth}
\centering
\includegraphics[width=1.\linewidth]{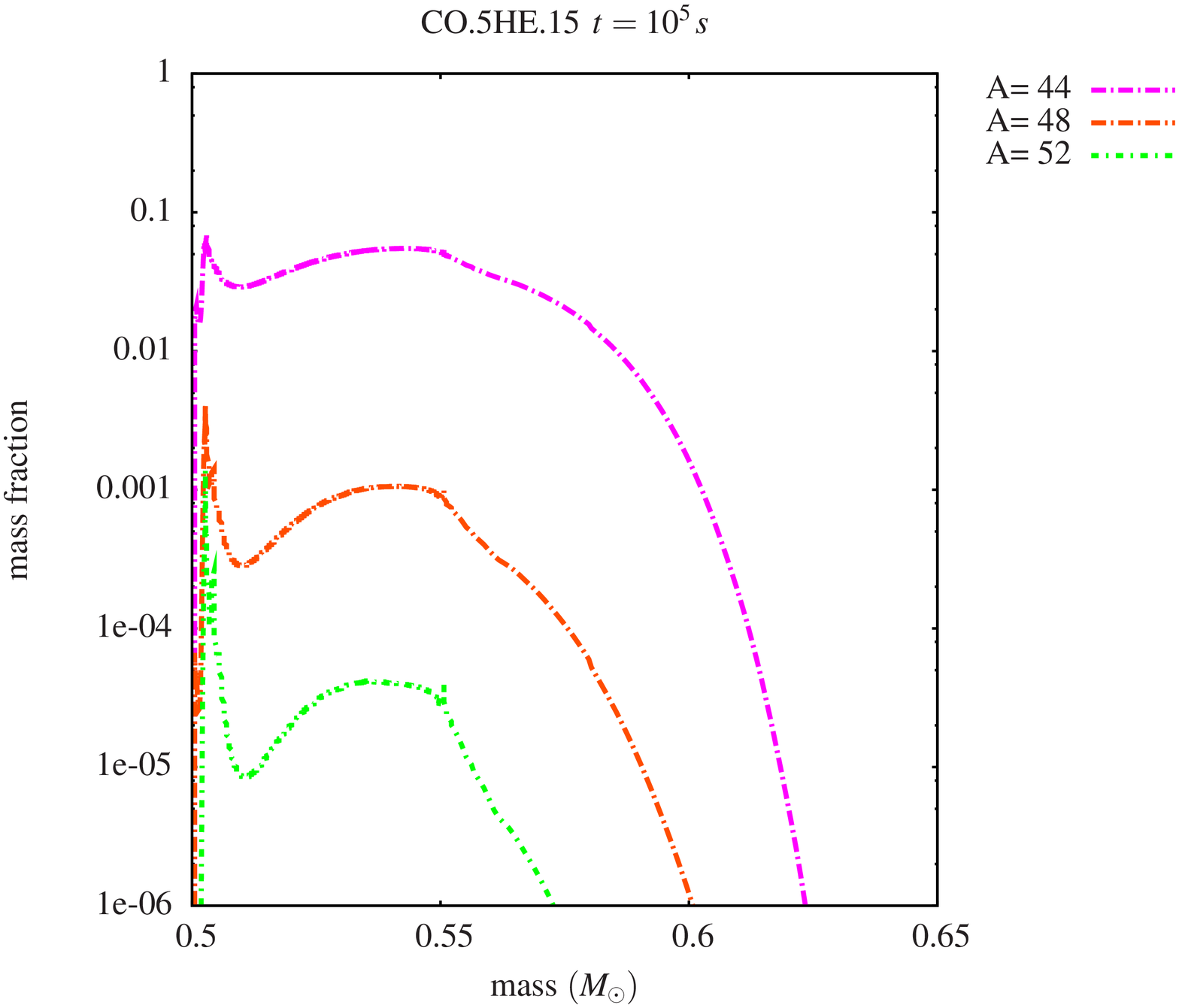}
\end{minipage}
\begin{minipage}[t]{0.5\linewidth}
\centering
\includegraphics[width=1.\linewidth]{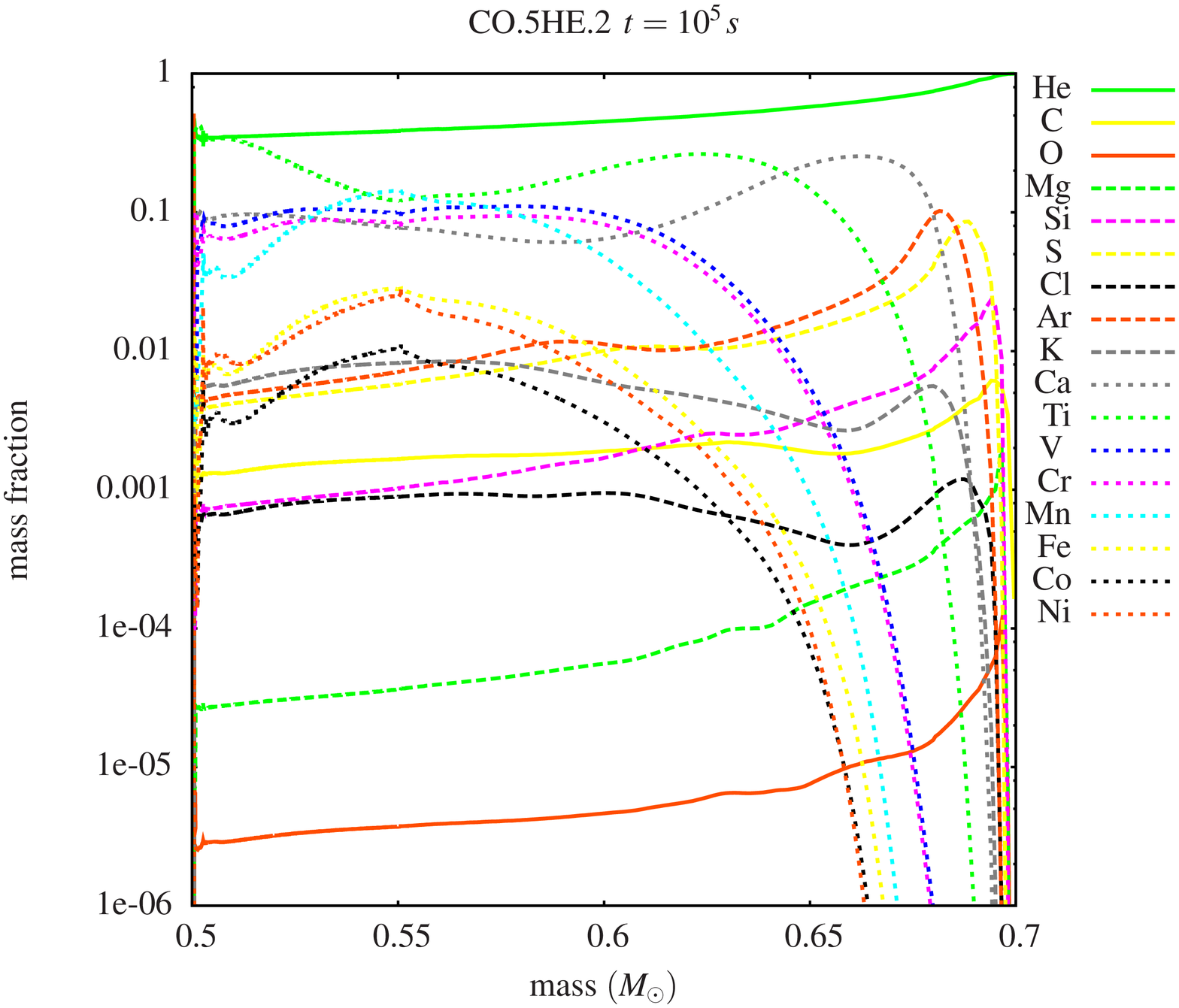}
\end{minipage}
\begin{minipage}[t]{0.5\linewidth}
\centering
\includegraphics[width=1.\linewidth]{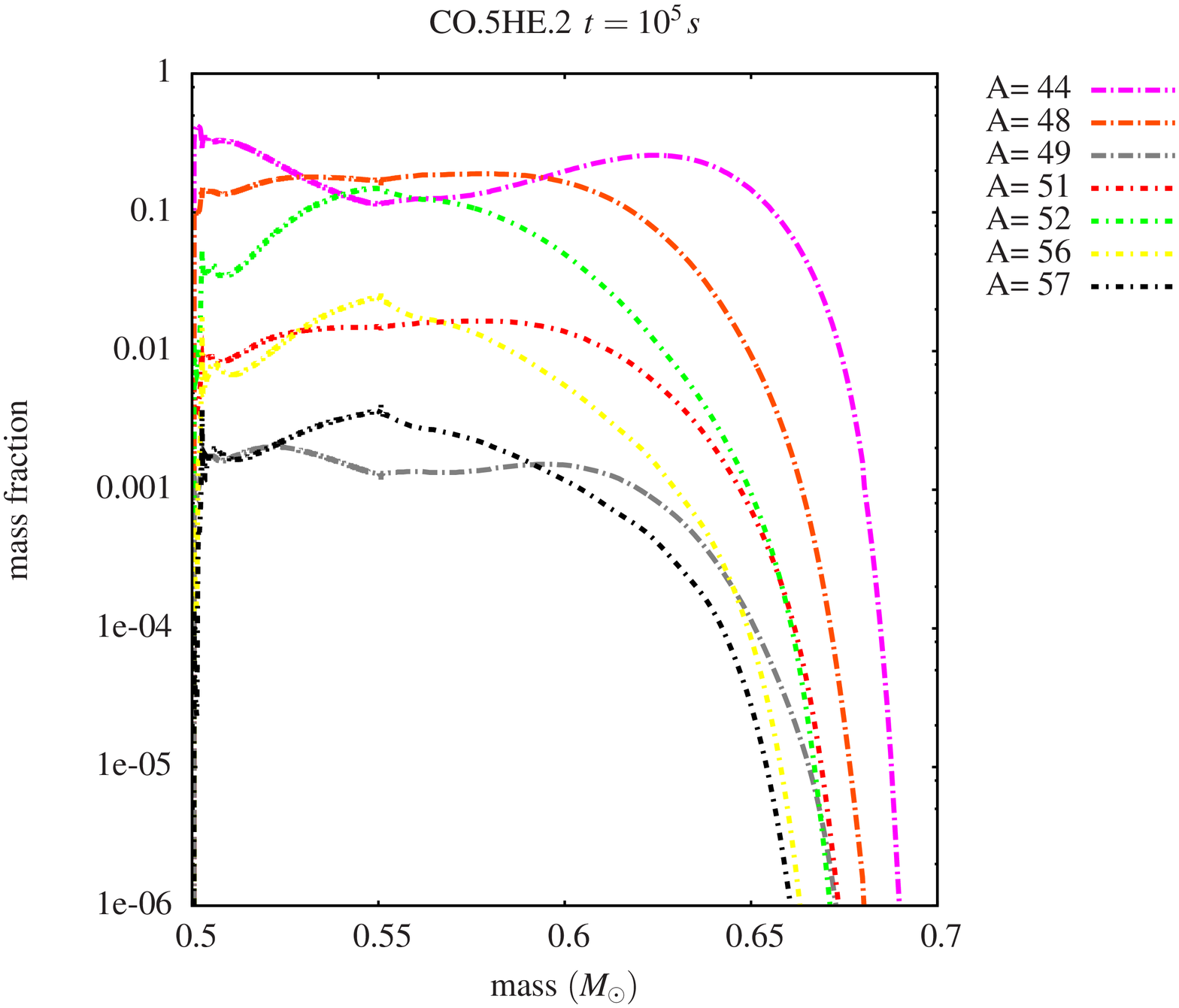}
\end{minipage}
\caption{The composition profile of our models at epoch $t=10^5\,{\rm s}$. For each model, the left panel sums up the species by elements; the right panel sums up the nuclei with equal atomic weight (marked on the plot), which form the decay chains followed in the radiative transfer calculations (see Sec.~\ref{radiative}). Compositions are showed only if the maximum abundance exceeds $10^{-3}$.}
  \label{fig:compos}
\end{figure*}

The compositions of our models are plotted in Fig.~\ref{fig:compos}. The plotted epoch is $10^5\,{\rm s}$, at which point the model is transferred to the radiative transfer calculation. For each model the figure shows (left panel) the composition by elements, and (right panel) the summed up composition of the species constituting the decay chains considered in the radiative transfer calculations (see Sec.~\ref{radiative}).

\begin{figure*}
\begin{minipage}[t]{0.5\linewidth}
\centering
\includegraphics[width=1.\linewidth]{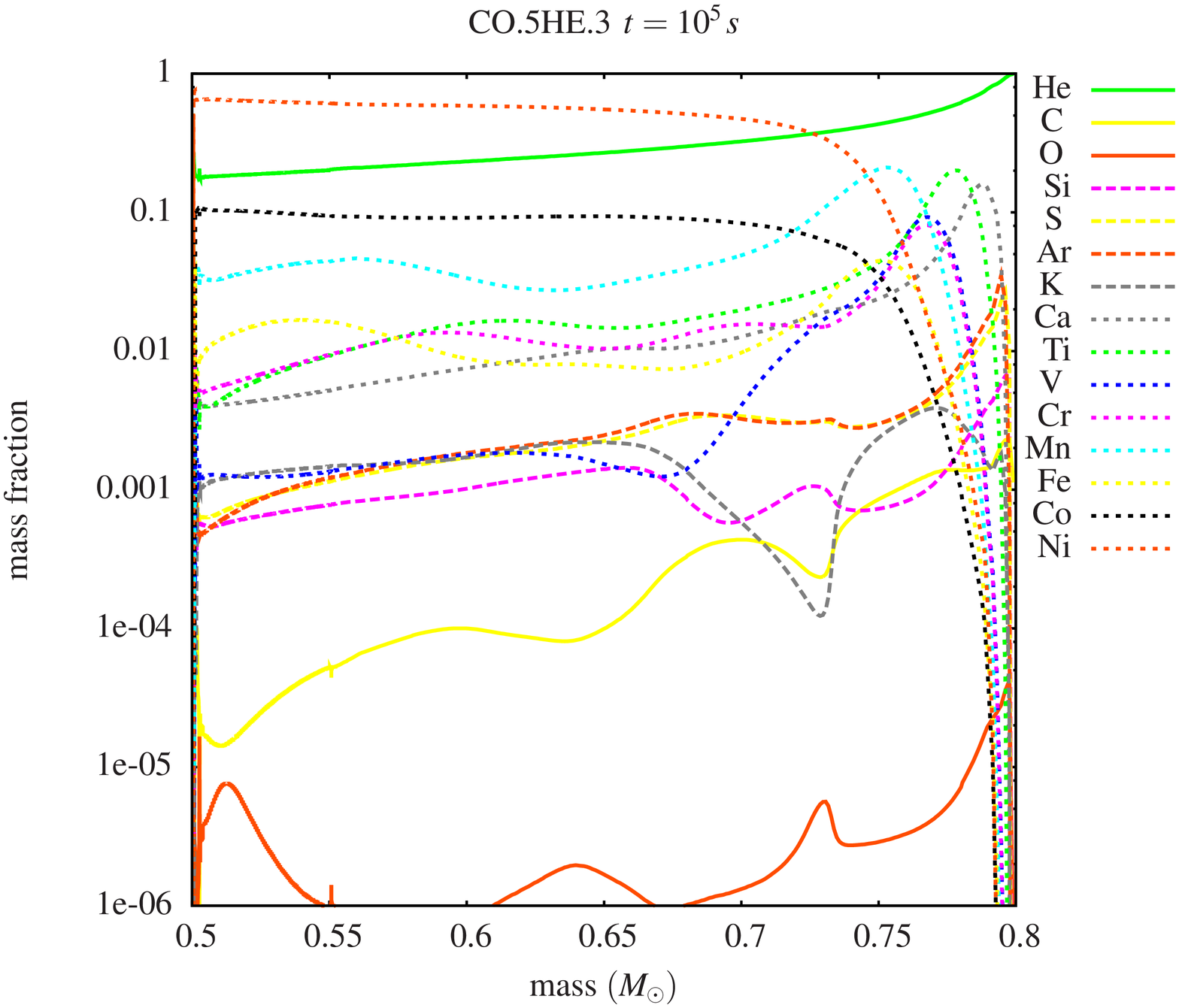}
\end{minipage}
\begin{minipage}[t]{0.5\linewidth}
\centering
\includegraphics[width=1.\linewidth]{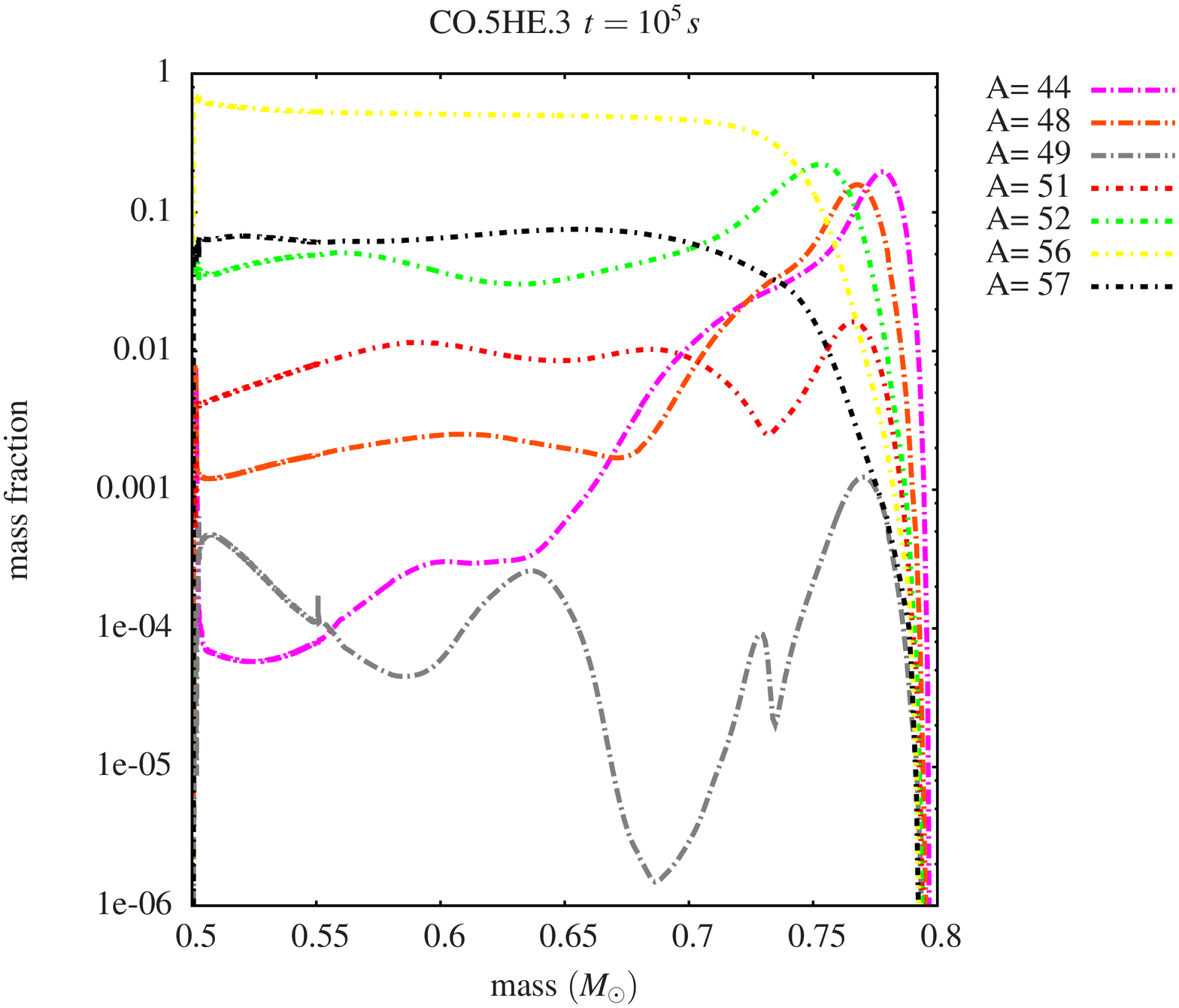}
\end{minipage}
\begin{minipage}[t]{0.5\linewidth}
\centering
\includegraphics[width=1.\linewidth]{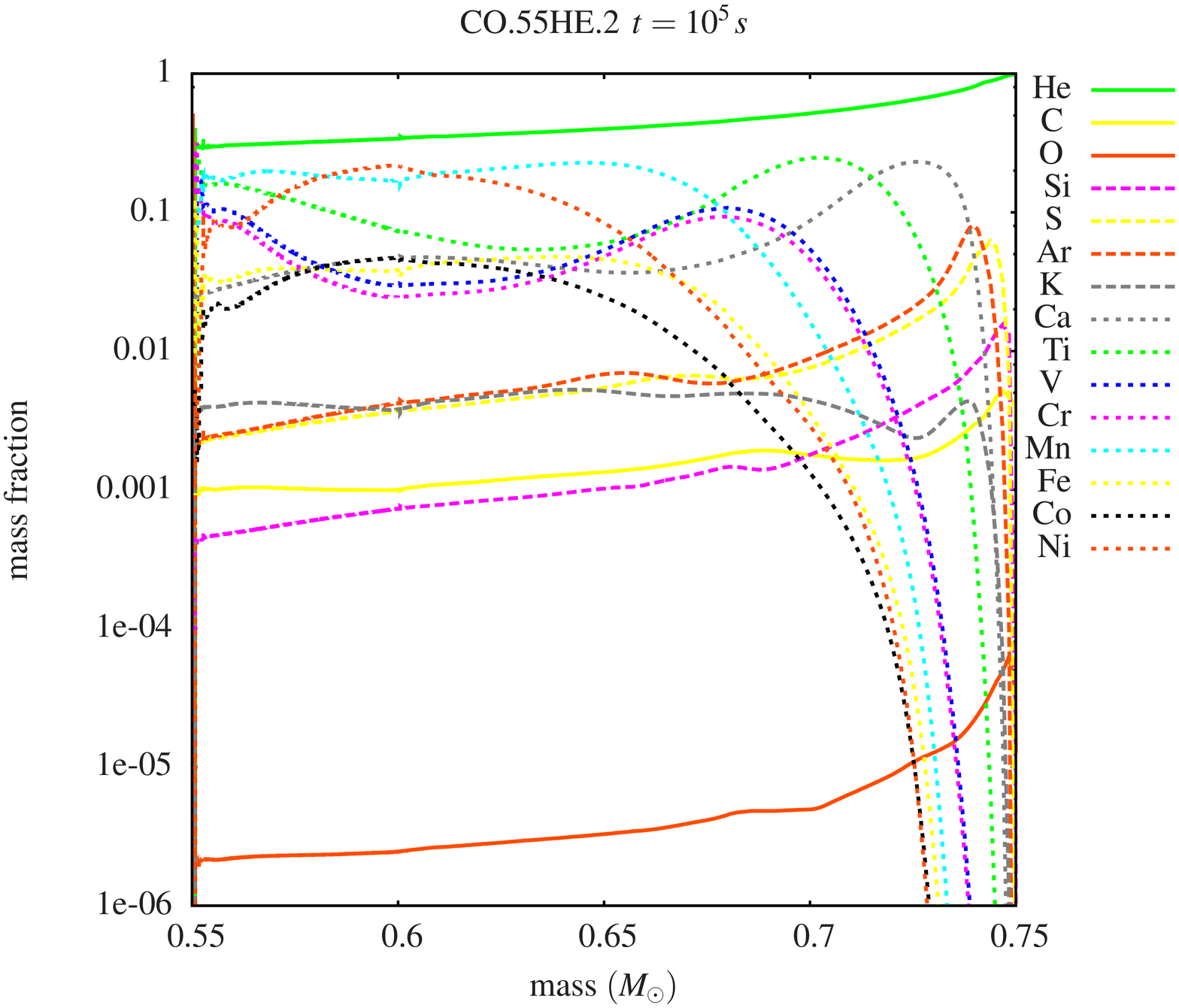}
\end{minipage}
\begin{minipage}[t]{0.5\linewidth}
\centering
\includegraphics[width=1.\linewidth]{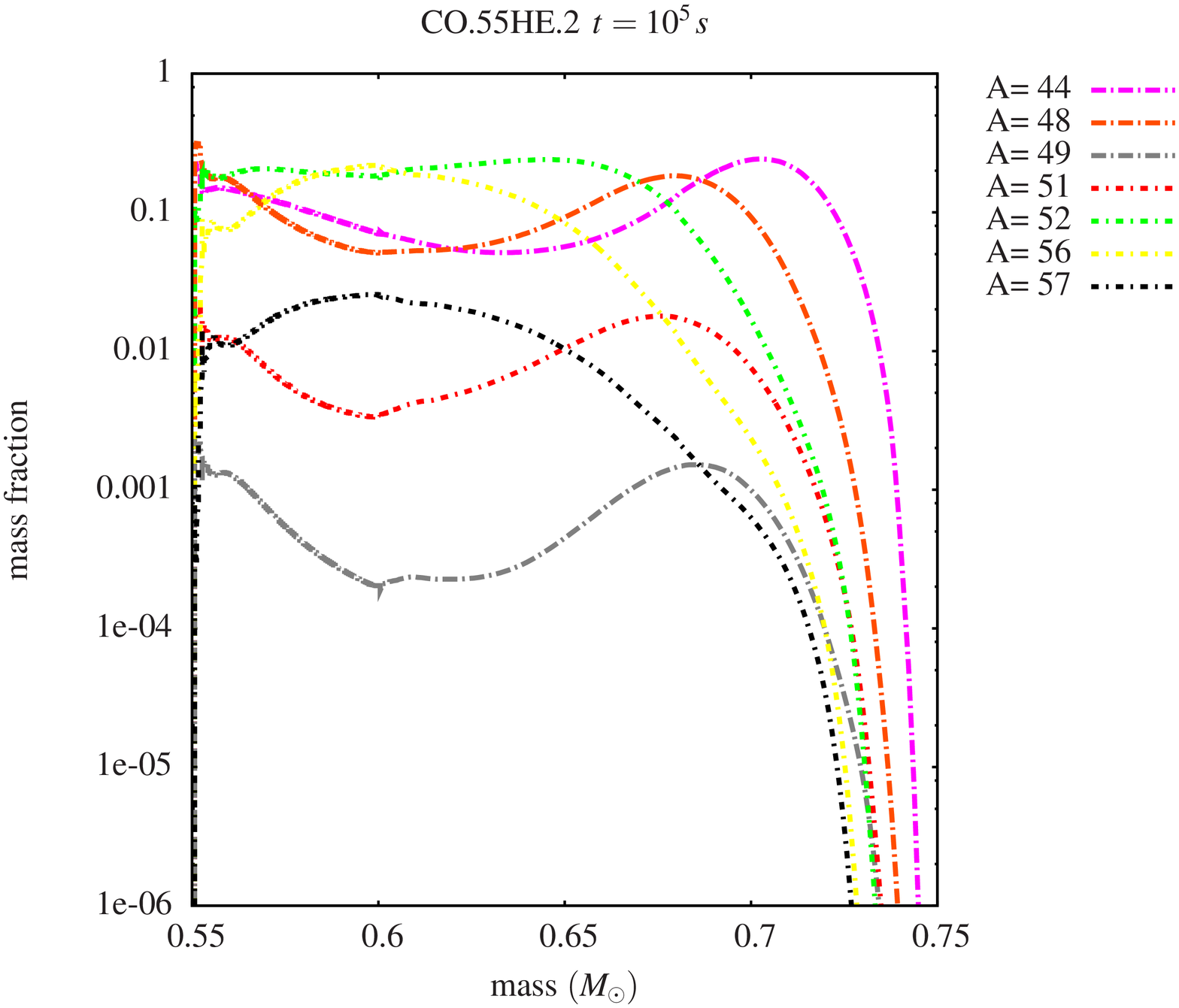}
\end{minipage}
\begin{minipage}[t]{0.5\linewidth}
\centering
\includegraphics[width=1.\linewidth]{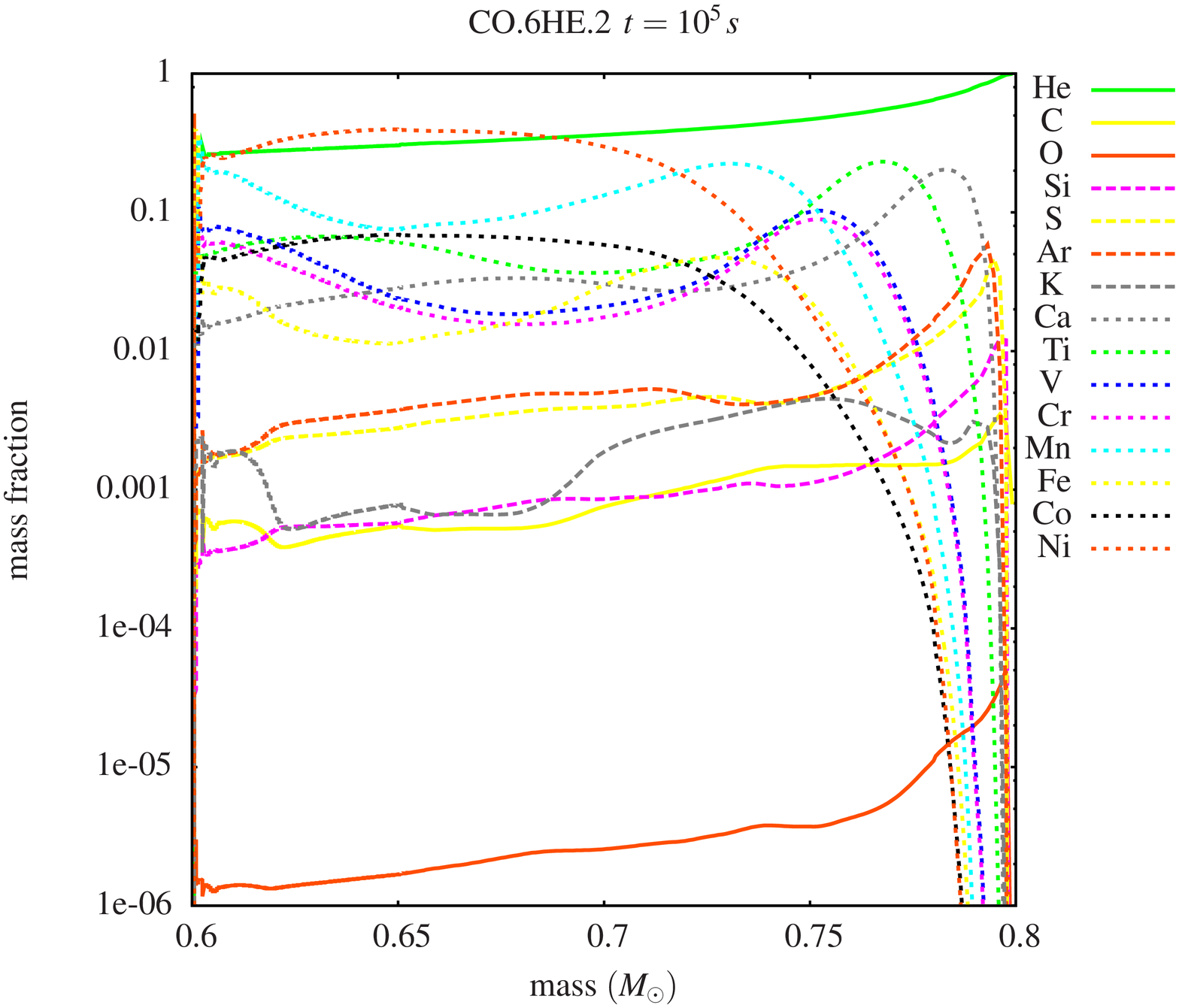}
\end{minipage}
\begin{minipage}[t]{0.5\linewidth}
\centering
\includegraphics[width=1.\linewidth]{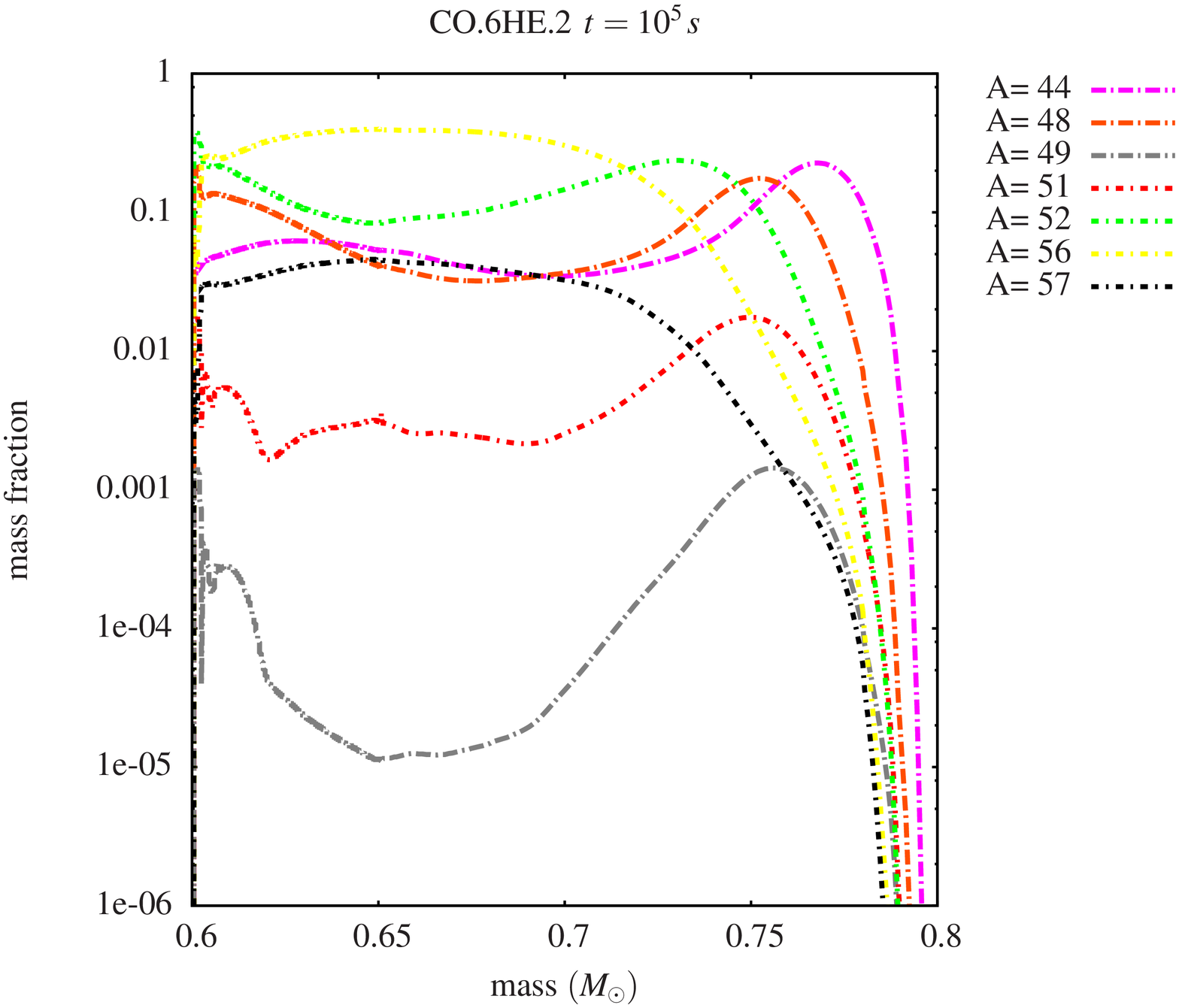}
\end{minipage}
\begin{center}
Fig.~\ref{fig:compos} (continued)
\end{center}
\end{figure*}

The main nucleosynthetic products from our simulations are summarized in Table~\ref{tab:exp_products}. \textbf{The main result of our simulations is the
sharp drop of radioactive nickel products for cores below $0.6\,\msun$}. Accordingly,
larger amounts of \iso{40}Ca, \iso{44}Ti and \iso{48}Cr are produced, with almost no silicon group elements.

In addition we note, that the remnant CO core is expected to undergo pulsations due to the inward moving shock wave initiated by the detonation being repeatedly reflected from the center and the outer boundary of the core, but this phenomenon is not adequately resolved in our present simulations.

\subsection{Effect of initial composition on nucleosynthesis}
\label{nucleosynthesis}

  A major drawback of pure helium detonations is the fact that
usually the burning goes all the way to \iso{56}Ni, and there is no
evidence for intermediate mass elements in the emerging
spectrum. One way to overcome this drawback, as we demonstrate
here, is to ignite the detonation at low densities. Once a pure
helium detonation is ignited at densities below approximately
$5\times 10^5\,{\rm g\,cm^{-3}}$, the burning rate is low enough so that the
composition, for the appropriate dynamic timescale, includes many
intermediate mass elements and almost no iron group elements (Table~\ref{tab:exp_products}, Fig.~\ref{fig:compos}).

  Another possible way to generate a composition of ejecta dominated
by intermediate mass elements is to ignite the detonation in a mixture
of helium and carbon (or CO). For the relevant temperatures we
obtain, the triple $\alpha$ reaction is slow compared to $\alpha$ capture on carbon.
Therefore, any pre existing \iso{12}C will tend to capture the free $\alpha$ particles.
It follows that, if there are sufficient carbon nuclei present at the onset of burning,
the end products can be calculated by a straight forward argument.

  The number N of available $\alpha$ particles per \iso{12}C is given by:
\begin{equation}\label{eq:alphas}
    N=(X_{He}/4)/(X_C/12)=3(X_{He}/X_C)
\end{equation}
 where $X_{He}$ and $X_C$ are the mass fractions of \iso{4}He and \iso{12}C respectively.

  Initially $X_C+X_{He}=1$, therefore the number N of
  available $\alpha$ particles a \iso{12}C nucleus will ultimately accumulate is
  $N=3(1-X_C)/X_C$.

  We assume here that $\alpha$ captures simply continue up the $\alpha$
  capture chain from \iso{12}C.

  In this way, for any initial pre existing carbon abundance $X_C$, we
can compute N and predict the end product. For initial carbon abundance $X_C$,
the predicted major final nucleus in the aftermath of the detonation is given in Table~\ref{tab:mix}.
Once the final nucleus is produced, there are no more free $\alpha$ particles available for
capture on heavy nuclei or for consumption by the triple $\alpha$ reaction.

  For a model in which we artificially assumed an initial carbon
abundance of $X_C=0.3$, there are seven free $\alpha$ particles for each
carbon atom and indeed the end product of the detonation is found to
be \iso{40}Ca, in accordance with the predictions given in Table~\ref{tab:mix}.

  A crucial issue for such a scenario is the origin of the pre existing carbon.
Assuming that the outburst occurs on a carbon WD that accretes helium,
carbon enrichment can take place if there is dredge up mixing at the bottom of
the envelope prior to the ignition of the detonation. Preliminary 1D models show
that the helium envelope is indeed unstable to convection about a day before the
runaway. We intend to examine this interesting possibility for the mixing process
by performing 1D and 2D simulations of the pre runaway evolution.

\section{Radiative transfer models}
 \label{radiative}
To connect the theoretical predictions from the explosion models to the
observed properties of SN~2005E, radiative transfer models which describe the
generation and propagation of light within the ejecta are needed.  Here we
investigate different aspects. First we discuss models of the bolometric light curve
describing the temporal evolution of the total light emission in ultraviolet,
optical and infrared ({\sc uvoir}) wavelength bands. Secondly, we computed a
sequence of synthetic spectra for the early epochs from the time of explosion
to about a week after maximum light.

\subsection{Bolometric light curve models}

The code we use to predict {\sc uvoir} light curves from the explosion models
provides a Monte-Carlo-based, time-dependent description of the propagation of
radiative energy packages through the ejecta. The procedure consists of a
simulation of the transport of energetic $\gamma$-photons released in the
radioactive decays within the ejecta and the transport of optical photons
generated after the interaction of the $\gamma$-photons with the matter.  The
procedure follows the description given in \citet{cappellaro97} and
\citet{mazzali00a}.

In addition to the decay of {\nifs} as the source of energetic $\gamma$-photons
we include also a number of other radioactive decay chains that are important
to describe the explosion models discussed here. The decays considered are:

$\begin{array}{lclcl}
\iso{57}{Ni}&\ra&\iso{57}{Co}&\ra&\iso{57}{Fe},\\
{\nifs}&\ra&{\cofs}&\ra&{\fefs},\\
\iso{52}{Fe}&\ra&\iso{52}{Mn}&\ra&\iso{52}{Cr},\\
\iso{51}{Mn}&\ra&\iso{51}{Cr}&\ra&\iso{51}{V},\\
\iso{49}{Cr}&\ra&\iso{49}{V}&\ra&\iso{49}{Ti},\\
\iso{48}{Cr}&\ra&\iso{48}{V}&\ra&\iso{48}{Ti},\\
\iso{44}{Ti}&\ra&\iso{44}{Sc}&\ra&\iso{44}{Ca}.
\end{array}$

For the transport of $\gamma$-photons
we adopt a constant gray opacity $\kappa_{\gamma}=0.027\,{\rm cm^{2}\,g^{-1}}$ and
assume that once a $\gamma$-package encounters an interaction it deposits all
its energy on the spot to generate a package of {\sc uvoir}-photons
\citep{swartz95}. The amount and rate of energy deposited in the ejecta is
determined following the method of \citet{ambwani88} and  \citet{lucy05}.
Some of the decays are accompanied by the emission of positrons which are
assumed to deposit their kinetic energy in situ.

The propagation of the resulting {\sc uvoir}-photons is modeled in a similar
Monte Carlo experiment also adopting a gray opacity $\kappa_{\rm uvoir}$ which is, however,
parameterized in terms of the abundance of Fe-group elements in the ejecta as
\begin{equation}
  \kappa_{\rm uvoir} = 0.25 X_{\rm Fe} + 0.025(1-X_{\rm Fe})\quad [{\rm cm}^{2}{\rm g}^{-1}].
  \label{eq:kappauvoir}
\end{equation}
This parametrization accounts for the much larger line opacity added by
the complex ions of Fe-group elements compared to lighter elements and has been
used in a number of previous studies \citep[e.g.,][]{mazzali01,mazzali06,sim07}.

A more detailed description of the modeling procedure will be published in an
accompanying paper (Sauer et al, in prep.) where we study more generally the
properties of the bolometric light curves from all explosion models described
in the beginning of this paper. Here we focus on the properties of the model
which provides the best fit to the observed light curve of SN~2005E.

\subsubsection{Results}

\begin{figure}[h]
  \plotone{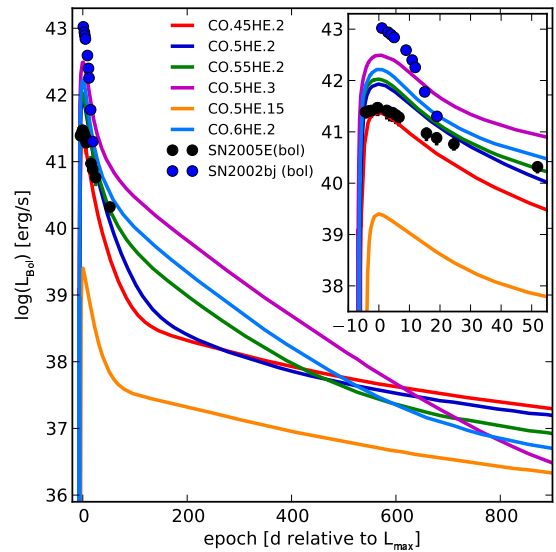}
  \caption{Bolometric light curves
    of all explosion models in comparison to the quasi-bolometric light curve
    points of the observed supernovae SN~2005E \citep[black circles,][]{perets10}, and SN~2002bj \citep[blue circles,][]{poznanski10}. The error bars
    indicate the range of the bolometric light curve points using different
    assumptions for unobserved wavelength bands (see text).  }
  \label{fig:lcallmod}
\end{figure}

We computed bolometric light curves for all explosion models in
Table~\ref{tab:models}. Fig.~\ref{fig:lcallmod} shows the resulting light curves
for all models in comparison to the quasi-bolometric light curve points for
SN~2005E (black symbols) and SN~2002bj (blue symbols). We derived the bolometric light curve points for SN~2005E from the
photometric data published in \citet{perets10} using the procedure developed in
\citet{valenti08a}.  In addition we made different assumptions to account for
the unobserved $U$ and $J\!H\!K$ bands indicated by the error bars on the data
points of SN~2005E.  For the lower limit we took only the observed
data from SN~2005E while for the upper limit we added the contribution of the
type Ic SN~2007gr \citep{hunter09}. The plot symbols correspond to the
intermediate light curve obtained from adding the $U$ and $J\!H\!K$ contribution
of the type Ia SN~2005cf \citep{pastorello07b}.  Also shown is the bolometric light curve
of SN~2002bj \citep{poznanski10}.

All models have a comparable rise time to maximum of about $7$ days.  Aside
from the absolute brightness at maximum light they show strong differences in
their late-time behavior.  The large range of different light curve properties
of those models is primarily due to the variation in the radioactive species
synthesized in the explosions which have different energy output and decay
timescales. In addition, the models have slightly different density structures
resulting from differences in mass and kinetic energy. The latter variation
affects the timescales for the diffusion of trapped photons from the ejecta and
the brightness at peak. A more in-depth comparison of the different light curve
models will be presented in a forthcoming publication. Here we focus merely
on the ability of the models to provide a good fit to the observed data of
SN~2005E. The best fit to the data points of SN~2005E is provided by the model
CO.45HE.2 which has a WD mass of $0.45\msun$ and a He-layer of $0.2\msun$. The
model is still somewhat faint at peak and clearly fades more rapidly after
maximum light than the observed light curve of SN~2005E. The rise time of the
model light curve to maximum light is $6.78\,$d. Unfortunately the pre-maximum data for SN~2005E is sparse and a good estimate of the rise time is not available; it seems that the rise time in our model is somewhat shorter than
the rise time of \tsim$11\,$d estimated for SN~2005E \citep{perets10}, but the errorbars on the latter are difficult to estimate with the available data.
None of our models light curves show similarities with the bright and very fast light curve
of SN~2002bj (\citealp{poznanski10}; or the other fast evolving SNe 1885A and 1939B; \citep{perets10c}, not shown). The latter bright fast evolving SNe are unlikely to be consistent with the models studied by us.

\begin{figure}[h]
  \plotone{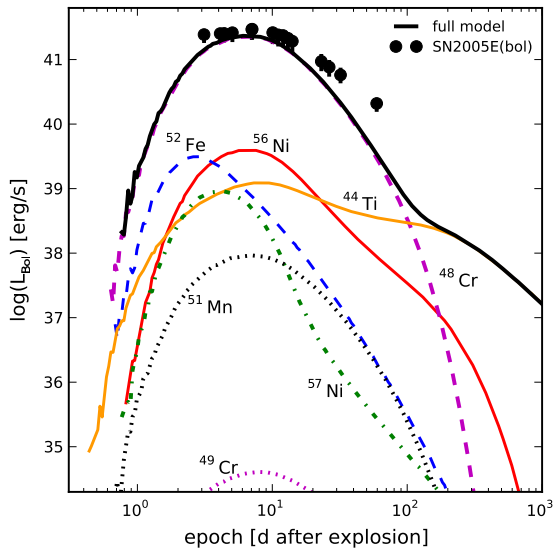}
  \caption{Model light curves from the explosion model CO.45He.2 (solid black
  line) in comparison to the quasi-bolometric light curve of SN~2005E (shifted to
  the same rise time as the model).  The other curves show the contribution of the
  different radioactive isotopes to the total light curve of this model, each
  labeled with the first isotope of the respective $\beta^{+}$-decay chain.}
  \label{fig:lccontrib}
\end{figure}

Figure~\ref{fig:lccontrib} shows the light curve of model CO.45HE.2 alone
(solid black line) in comparison to the observed data points (shifted
to have the same rise time as the model). In addition, the
contribution of the different $\beta^{+}$ decay chains to the total light
curves are shown, labeled with the first isotope of each chain. In this
particular model \iso{44}{Ti} is actually the most abundant radioactive species
by mass, however, due to its long half life of $60$ years, this decay only
dominates the light curve at late times after \tsim$200\,$d. The dominant
contributor to the light emission around peak is the decay chain  \iso{48}{Cr
}$\ra$\iso{48}{V}\ra\iso{48}{Ti}. The time scale of this decay is dominated by
the half life of the second decay of $15.97\,$d. The contribution of the
\iso{56}{Ni}/\iso{56}{Co} decays, the most important source of radiative energy
in normal, radioactively powered supernova light curves, is two orders of
magnitude less than the \iso{48}{Cr}/\iso{48}{V} contribution in this model.

\begin{figure}[h]
  \plotone{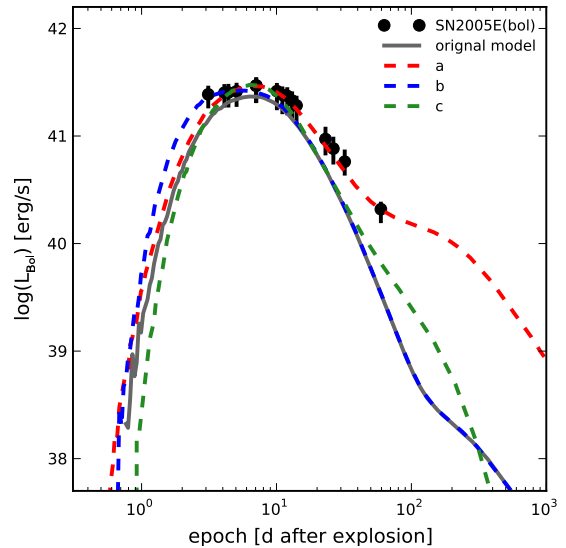}
  \caption{Tests if variation of
  the relative contributions of the radioactive elements can explain the
  observed shape of the light curve of SN~2005E. In all models  only the
  relative contributions of the individual light curves shown in
  Fig.~\ref{fig:lccontrib} are varied. The  solid gray line refers to the
  original model, model {\it a} was obtained by increasing the \iso{44}{Ti}
  contribution by a factor of $50$. For model {\it b} \iso{52}{Fe} was enhanced
  by a factor of $30$, model {\it c} was obtained by only using the
  \iso{56}{Ni} contribution times a factor of $75$ while ignoring all other
  isotopes. }
  \label{fig:tweaklc}
\end{figure}

We also tried to check whether the observed light curve shape can be
matched in a better way with slightly different contributions from the various radioactive isotopes.
To test this possibility
we varied the relative contributions of the different decay chains to the total
light curve. In Fig.~\ref{fig:tweaklc} we show three different models in
addition to the original model (solid gray line). In all models the density
structure is kept the same, only the relative contributions of the individual
light curves shown in Fig.~\ref{fig:lccontrib} are varied.

Curiously, if one could increase the energy output from the \iso{44}{Ti} decay
by a factor of \tsim$50$ this model would provide an excellent fit also to the
late-time light curve of SN~2005E (model {\it a} in Fig.~\ref{fig:tweaklc}). However,
because \iso{44}{Ti} is already the most abundant Fe-group element in this
model, this increased power cannot be accomplished without changing the mass
and density structure of the entire model, which would inevitably lead to a
completely different light curve behavior.  Nevertheless, this may indicate
that the light emission of SN~2005E is dominated by the \iso{44}{Ti} decay
already at times shortly after maximum light. A more conclusive assessment
would require observations out to much later times of several hundred days
after the explosion where one can safely assume that most of the trapped
radiation has been released from the ejecta and the short-lived isotopes do not
contribute significantly to the light emission anymore.

The pre-maximum data of SN~2005E also indicate that this supernova brightened
earlier than the model light curve. A larger contribution of the \iso{52}{Fe}
light curve by a factor of \tsim$30$ could explain the early brightening (model
{\it b} in Fig.~\ref{fig:tweaklc}).  However, given the sparse data available for
SN~2005E a larger ejecta mass leading to a wider light curve cannot be excluded.

\citet{perets10} estimate an ejected mass of $0.275\msun$ and $0.003\msun$ of
{\nifs} from nebular models of SN~2005E which, however, do not consider any
other sources of radiative energy.  Model CO.45HE.2 contains
$1.1\times10^{-4}\msun$ of {\nifs}.  To reproduce the peak of the light curve
with {\nifs} alone, a factor of $\tsim70$ more mass  of this isotope (i.e.,
\tsim$7.8\times10^{-3}\msun$) is needed (model {\it c} in Fig.~\ref{fig:tweaklc}).
However, with the given density structure {\nifs} alone cannot reproduce the
decline after maximum. The model light curves decay faster than the observation.

In summary, the decline of the post-maximum light curve of SN~2005E indicates
that the ejected mass in SN~2005E might be somewhat larger than proposed by the
explosion models. The \iso{44}{Ti} decay in the model can explain the shape but
not the absolute brightness of the late time decline.  The pre-maximum light
curve also hints towards a larger ejecta mass, some contribution to the early
brightening might be explained by the contribution of the \iso{52}{Fe} decay
chain.

\subsection{Spectral models}

We used our Monte Carlo spectral synthesis code to derive a sequence of
spectral models from the explosion models. The code is based on the description
by \citet{mazzali93b,lucy99}, and \citet{stehle05} and has been successfully used
in the past for the efficient interpretation of observed supernova spectra
\citep[e.g.,][]{mazzali06b,mazzali08,sauer08}.
The numerical methods are described in more detail in the aforementioned
publications, here we discuss only the most important assumptions as far as
they are important for the models.

The code computes a stationary solution for the radiative transfer through the
supernova ejecta using a Monte Carlo method with an approximate non-LTE
description for the atomic level populations.  The underlying density structure
is taken from the hydrodynamic model for explosion and is expanded homologously
according to the time after explosion for each model, assuming that the
radiation does not alter the hydrodynamic structure of the explosion.  The
radius-dependent composition  is extracted from the nucleosynthesis results of
the explosion model. For radioactive species with appreciable abundance and
relevant decay time scales, the conversion to the respective daughter elements
is taken into account according to the epoch. Specifically, the decay chains
treated explicitly are:

$\begin{array}{lclcl}
\iso{57}{Ni}&\ra&\iso{57}{Co}&\ra&\iso{57}{Fe},\\
{\nifs}&\ra&{\cofs}&\ra&{\fefs},\\
\iso{52}{Fe}&\ra&\iso{52}{Mn}&\ra&\iso{52}{Cr},\\
\iso{48}{Cr}&\ra&\iso{48}{V}&\ra&\iso{48}{Ti}.
\end{array}$

The solution of the radiative transfer assumes a Schuster-Schwarzschild
situation imposing a sharp inner boundary where all radiation is emitted as a
black-body. This implies that the $\gamma$-photons and positrons from
radioactive decays below this inner boundary are assumed to thermalize, the
energy deposition above the inner boundary is not taken into account.  In
supernova ejecta it is generally difficult to make a clear choice of where to
place the inner boundary because the density structure of unbound ejecta tends
to be very shallow such that the location where the ejecta become optically
thick varies strongly with wavelength \citep[e.g.,][]{sauer06}. For the models
discussed here we chose this location iteratively based on the dilution of the
radiation field such that the dilution factor $W$  becomes close to $0.5$ (see
\citealt{mazzali93b} for a discussion of how $W$ is determined in our model). With
that choice $v_{0}$ corresponds to the location where the radiation field
becomes roughly isotropic.

The approximation of stationarity also implies that we cannot determine the
luminosity of the model self-consistently because the emitted radiative energy at a
given time has contributions from both the directly deposited energy from
radioactive decays and the radiation originating from earlier decays which is
stored in the optically thick ejecta and diffuses out as the ejecta expand.
Therefore, we use the bolometric luminosity at a given epoch from the models of
the bolometric light curves discussed in the previous section.

Once the ejecta start to become diluted enough that they are transparent over a
wide range of the spectrum, the lower-boundary approximation will fail. For the
explosion models discussed here this generally happens already fairly early
at about 10 days after maximum because the ejected mass of all models is low.
In contrast to the light curve models which do not depend on the photospheric
assumption, we cannot compute meaningful spectra beyond the epoch where
that approximation breaks down.

Another complication involves the non-thermal excitation of \ion{He}{1}.
\ion{He}{1} has high-lying energy levels such that the ejecta temperatures are
too low to populate those levels thermally to give rise to visible absorption
lines in the spectrum. The \ion{He}{1} absorption features seen in SN~Ib
(hydrogen-deficient supernovae with visible helium lines, see
\citealt{filippenko97} for a detailed discussion of different types of
supernova spectra) are caused by non-thermal excitations from fast electrons
which result from Compton-scatterings of the $\gamma$-photons released by
radioactive decays in the ejecta \citep[e.g.][]{lucy91,kozma92,mazzali98a}.

The spectral code we use in this study does not include those non-thermal
excitations and ionizations self-consistently. Therefore, the models will not
show significant \ion{He}{1} absorption features even though the explosion models
predict that the He-rich material is well mixed into the zones that have a high
abundance of radioactive isotopes. Nevertheless, assuming that the opacity
contributed by the relatively few \ion{He}{1} lines will not substantially
affect the model, the other aspects of the synthetic spectra should predict the
observable characteristics of the explosion models reasonably well.

\subsubsection{Results}

\begin{figure}[h]
  \plotone{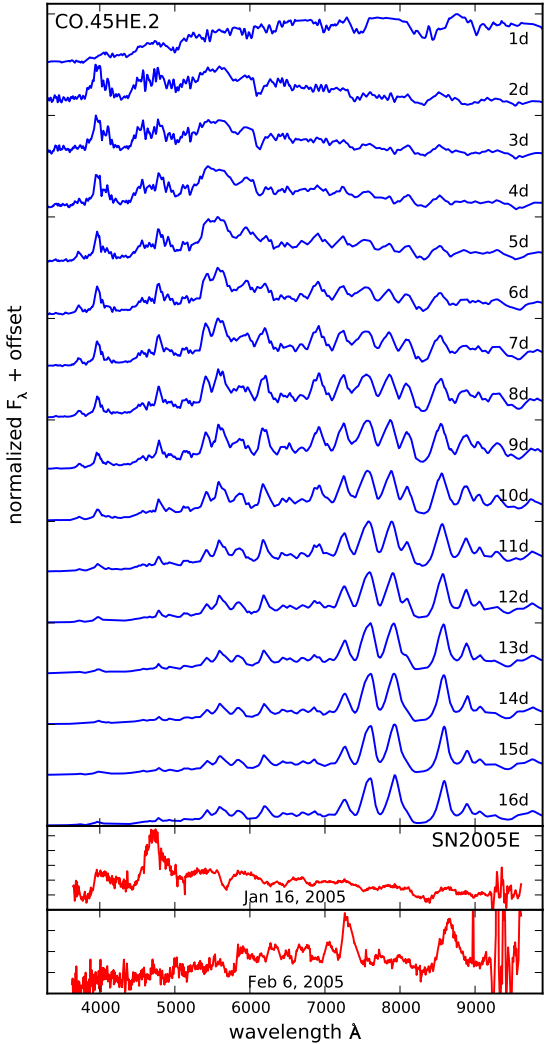}
  \caption{Series of synthetic spectra for the explosion model CO.45HE.2 (upper
    panel). The epoch indicated at each spectrum is given relative to the
    explosion date. The absolute flux scale has been normalized to the peak of
    each spectrum to allow for a comparison of the spectral shapes. The (bolometric)
    maximum of this model occurred  $6.78\,$d
    after explosion.  The lower panels show the observed spectra of SN~2005E
    \tsim3d before and \tsim18\,d after maximum light.}
  \label{fig:specseries}
\end{figure}

\begin{figure}[h]
  \plotone{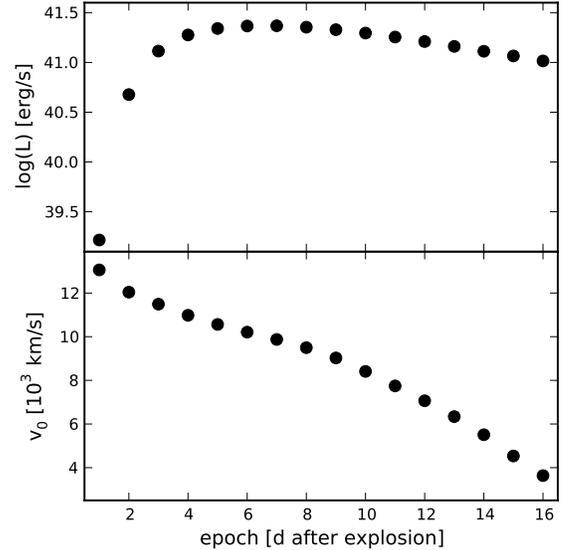}
  \caption{Input luminosity (upper panel) and velocity of the inner boundary
  (lower panel) for the spectral models as a function of time after explosion.}
  \label{fig:vphl}
\end{figure}

Fig.~\ref{fig:specseries} shows a series of synthetic spectra from the
explosion model CO.45HE.2 between day 1 and 16 after explosion. The flux of
each spectrum in this plot has been normalized with respect to the maximum flux
to make the differences in the spectral shape visible.  The lower two panels
show the two spectra of SN~2005E available during the photospheric phase. The
Jan 16 spectrum corresponds to an epoch of \tsim$3\,$d before maximum light,
the Feb 6 spectrum was taken \tsim$18\,$d after maximum.  The input values for
the luminosity and the inner boundary velocity $v_{0}$ for all spectral models
in this series are shown in Fig.~\ref{fig:vphl}.

\begin{figure}[h]
  \plotone{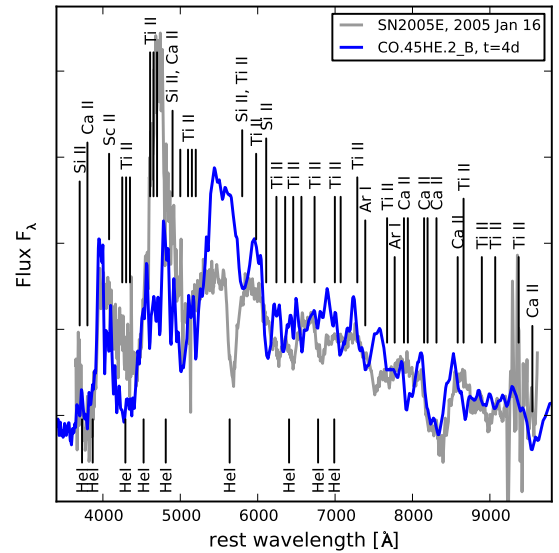}
  \caption{Comparison of the synthetic spectrum from the model CO.45He.2 four
    days after explosion (corresponding to \tsim3 days before maximum light) to
    the observed spectrum of SN~2005E. The flux of the model has been multiplied
    by a constant to match the observed brightness of SN~2005E at this epoch.
    The labels above the spectra indicate the
    ions that form the main contribution to the respective absorption feature.
    The labels below the spectra indicate the positions where the stronger
    \ion{He}{1} lines are expected in the model spectrum if the model would treat
    non-thermal excitation of \ion{He}{1} (see text).}
  \label{fig:compare05e}
\end{figure}

The most prominent ion dominating most of the spectral shape at all times is
\ion{Ti}{2}. Also visible are the characteristic absorption features from
\ion{Ca}{2} at \tsim$3900\,${\AA} and around $8300\,${\AA}.  The first spectrum
in this series is fairly red because of the low luminosity on day 1. Day 2
shows then a much bluer spectrum which gradually becomes redder as time goes
on. All spectra show only very little flux blue of the \ion{Ca}{2} absorption
at \tsim$3900\,${\AA}. In this region the flux is effectively blocked by a
dense forest of lines from Fe-group elements, mostly \ion{Ti}{2} (cf. Fig \ref{fig:compare05e}).  Comparing
the overall shape of the model spectra to the two observed spectra one notices
that the models show most of the observed features, however, the strength of
absorptions and emission peaks are not well reproduced.  The spectra before
maximum light show a prominent peak at \tsim$4000\,${\AA} which is not as
prominent in the observed pre-maximum spectrum. In contrast, the observed
spectrum shows a prominent feature at \tsim$4800\,${\AA} which is partially
suppressed by absorption features in the model series.

In the post-maximum phases the model spectra evolve
faster than the observed spectrum would suggest. This
is consistent with the light curve model not
reproducing the observed late-time behavior of SN~2005E.
Note that the epoch of the Feb 6 spectrum relative to the observed maximum of SN~2005E is later than the relative epoch of the last model spectrum in the series. The model spectra after maximum light show strong re-emission features between $7500$ and $9000\,${\AA}.
Indication for those features are present in the observed spectrum, however at very different relative strengths. The strong features in the model spectra originate from strong absorptions and the respective re-emission peaks from \ion{Ti}{2} lines.

Fig.~\ref{fig:compare05e} shows a detailed comparison of the model spectrum for
day 3 and the observed pre-maximum spectrum of SN~2005E. The flux of the model
spectrum has been scaled by an arbitrary factor to match the luminosity of the
observed spectrum by eye. In this figure the most important ions contributing
to observed spectral features are labeled above the spectra. In particular
\ion{Ti}{2} lines also contribute to most other features in the spectrum.
However, for clarity not all are labeled explicitly. Below the spectra
the positions are indicated where one would expect the strongest \ion{He}{1}
lines to show up if the model included non-thermal excitations of this ion.

The overall shape of the spectrum is reproduced reasonably well by the model. The line
velocities and widths are overall correct indicating that the ejecta velocities
observed in SN~2005E can be explained by the explosion model.
The strong peak at \tsim$4700\,${\AA} seen in the observed spectrum is blocked
by a number of strong \ion{Ti}{2} lines in the model. Consequently, we see a
strong re-emission peak redwards at \tsim$5500\,${\AA}. The model shows
\ion{Si}{2} lines at $4880$, $5800$ and $6100\,${\AA}, which are not or only
very weakly present in the observed spectrum. Especially the absence of the
strong \ion{Si}{2} 6347, 6371 absorption in the observation indicates that the
ejecta of SN~2005E contain only little Si. The total mass of Si in the ejecta
of model CO.45HE.2 is $10^{-3}\msun$.

In summary, it appears that the composition of the explosion model can
reproduce the observed spectrum, perhaps the Ti content is somewhat too high in
the model, however the features from intermediate mass elements such as S
and Si are also stronger in the model than in the observation. A caveat is
that the radioactive isotopes are present in the ejecta at almost all
velocities. The effects of non-thermal ionizations and excitations caused by the
decays are not accounted for in the radiative transfer model.


\section{Conclusions}

Limited to spherical symmetry, we explored the consequences of helium detonations
in helium layers accreted on CO cores for a range of core masses. Recent studies by \cite{shen10} explored
similar scenarios. Their models, however, focused on CO cores of $\geq 0.6\,\msun$, while
we studied models with CO cores of  $\leq 0.6\,\msun$, and extended the range
of masses to the lowest reasonable limit of $0.45\,\msun$.
In all cases we assumed that carbon is not ignited.
Our study and that of \cite{shen10} are therefore complementary. We presented here the scenario
of a CO core of $0.6\,\msun$ with $0.2\,\msun$ of helium layer. This scenario
is comparable to the lowest mass cores scenarios explored by \cite{shen10}. Given the
different methods and simulation codes used, our results and those of \cite{shen10} are
in good agreement, showing quite similar light curves and nucleosynthesis products.

Our results show that below a CO core mass of $0.6\,\msun$, the abundance of
\iso{56}Ni, produced by the detonation, drops rapidly, while those of \iso{40}Ca, \iso{44}Ti, and \iso{48}Cr
grow rapidly to a few percent.
Moreover, the fraction of unburnt helium also grows
with decreasing CO core mass. These findings are encouraging for the discussed scenario
to be a viable model for SN~2005E-like objects, and suggest that the latter, low mass CO core scenarios on which we focused are the more likely progenitors of these explosions.
It is important to note that, in a narrow mass range of models, a highly diverse outcome is expected, both in terms of light curves and nucleosynthesis. Hence, one cannot expect an accurate fit of a model to a specific observed supernova.
Nevertheless, more detailed calculations of the
spectra, including the helium lines and late nebular spectra are yet to be done.

Besides the spectrum, the main remaining question concerns the multi dimensional aspect
of the problem, which has a strong impact on whether a consecutive CO detonation
may follow the helium detonation or not. We speculate that in the 2D case and with low CO core masses
($0.5\,\msun$ and under), the density of the core around the possible ignition sites will be too low for CO ignition.
The late evolution to thermonuclear runaway and the onset of helium detonation are also issues which
require at least two dimensional tools. This is due to the convective nature of the reactive
flow at the base of the helium layer, prior to runaway. Dredge up of heavier nuclei into the helium layer
could in principle alter the conditions for a transition from quasi-static burning to a detonation.

In terms of nucleosynthetic products, the large abundances of \iso{40}Ca and \iso{44}Ti produced in our detailed explosion models confirm similar results obtained with the simplified one-zone nucleosynthetic calculation discussed in \cite{perets10}. As discussed there, our detailed model results could therefore have important implications for the calcium enrichment of the interstellar medium as well as the the production of positrons and the production of the 511 keV annihilation line.

Finally, if SN~2005E type objects are produced by helium detonations on small CO cores, there
should also be a subclass of objects formed by helium detonations on larger cores. In that
class the spectrum should be dominated by lines of radioactive nickel, and the probable ignition
of the CO core should produce much more energetic and metal rich objects. We therefore encourage
observers to search for such supernovae in the near future.

\acknowledgments  Thanks to Stefano Valenti for computing the quasi-bolometric
light curve of SN~2005E from the observed data. We are grateful for Stuart Sim
and Markus Kromer at the Max-Planck Institute for Astrophysics for computing a
radiative transfer model to cross-check our model results.  Nuclear half lives
and decay energies used in the light curve model calculations have been
compiled using NuDat 2.5 ({\tt http://www.nndc.bnl.gov/nudat2}) and the
Web-based Table of Nuclides of the Korea Atomic Energy Research Institute ({\tt
http://atom.kaeri.re.kr}). Joint research by A.G. and P.A.M. is supported by a Weizmann-Minerva grant. The work of A.G. is also supported by the Israeli Science Foundation, an EU FP7 Marie Curie IRG fellowship, and a research grant from the Peter and Patricia Gruber Awards. This work is supported in part at the Argonne National Laboratory
 by the DOE under contract No. DE-AC02-06CH11357, and at the
 University of Chicago by the NSF under grant PHY 08-22648 for the
 Physics Frontier Center ``Joint Institute for Nuclear Astrophysics''
 (JINA).

\newpage

\begin{deluxetable*}{lcccccccc}
\tablewidth{15cm}
\tabletypesize{\scriptsize}
\tablecaption{Parameters of the Simulated Initial Configurations
\label{tab_bw}}
\tablehead{
\colhead{Model}&
\colhead{$M_{CO}$}&
\colhead{$M_{He}$}&
\colhead{$\rho_{6c}$}&
\colhead{$T_{7c}$}&
\colhead{$\rho_{6He}$}&
\colhead{$T_{7He}$}&
}
\startdata
 CO.45HE.2 &  0.45  &   0.2    &   3.81   &  1. &  0.543  &  20.   \\
\\
 CO.5HE.15 &  0.5   &   0.15   &   3.92   &  1. &  0.402  &  20.   \\
\\
 CO.5HE.2 &  0.5   &   0.2    &   5.06   &  1. &  0.678  &  20.   \\
\\
 CO.5HE.3 &  0.5   &   0.3    &   8.50   &  1. &  1.391  &  20.   \\
\\
 CO.55HE.2 &  0.55  &   0.2    &   6.72   &  1. &  0.845  &  20.   \\
\\
 CO.6HE.2 &  0.6   &   0.2    &   8.81   &  1. &  1.032  &  20.   \\
\enddata
\tablecomments{mass units=\m, $\rho_{6c}$ is the central density in units of $10^6 g\,cm^{-3}$,
$\rho_{6He}$ is the density at the base of the helium shell in same units, $T_{7c}$ and $T_{7He}$ are
the corresponding temperatures in units of $10^7K$.}
\label{tab:models}
\end{deluxetable*}

\begin{deluxetable*}{lcccccc}
\tablewidth{15cm}
\tabletypesize{\scriptsize}
\tablecaption{Explosion Energy and Nucleosynthesis Products
\label{tab:exp_products}}
\tablehead{
\colhead{Isotope}&
\colhead{CO.45HE.2}&
\colhead{CO.5HE.15}&
\colhead{ CO.5HE.2}&
\colhead{CO.5HE.3}&
\colhead{CO.55HE.2}&
\colhead{CO.6HE.2}
}
\startdata
Ek &   0.178 & 0.096 & 0.201 & 0.460 & 0.226 & 0.242 \\
\iso{4}He & 1.1E-01 & 9.8E-02 & 1.0E-01 & 9.7E-02 & 9.3E-02 & 8.3E-02 \\
\iso{20}Ne & 9.8E-06 & 1.2E-05 & 6.5E-06 & 3.9E-06 & 4.7E-06 & 3.7E-06 \\
\iso{23}Na & 1.2E-09 & 2.0E-09 & 6.0E-10 & 1.4E-09 & 3.0E-10 & 1.6E-10 \\
\iso{24}Mg & 4.4E-05 & 6.2E-05 & 2.7E-05 & 1.3E-05 & 1.8E-05 & 1.4E-05 \\
\iso{27}Al & 9.9E-07 & 1.1E-06 & 7.0E-07 & 3.1E-07 & 4.5E-07 & 4.0E-07 \\
\iso{28}Si & 1.0E-03 & 1.5E-03 & 5.8E-04 & 1.7E-04 & 3.8E-04 & 2.7E-04 \\
\iso{31}P & 1.8E-05 & 2.2E-05 & 1.2E-05 & 2.6E-05 & 8.3E-06 & 6.6E-06 \\
\iso{32}S & 4.6E-03 & 6.9E-03 & 2.7E-03 & 5.5E-04 & 1.7E-03 & 1.1E-03 \\
\iso{35}Cl & 1.4E-04 & 1.3E-04 & 1.5E-04 & 1.3E-04 & 1.2E-04 & 6.6E-05 \\
\iso{36}Ar & 5.5E-03 & 9.1E-03 & 3.2E-03 & 6.0E-04 & 1.9E-03 & 1.4E-03 \\
\iso{39}K & 9.4E-04 & 6.8E-04 & 1.1E-03 & 4.9E-04 & 8.2E-04 & 4.1E-04 \\
\iso{40}Ca & 3.4E-02 & 2.9E-02 & 2.1E-02 & 5.5E-03 & 1.3E-02 & 9.0E-03 \\
\iso{41}Ca & 3.3E-06 & 1.8E-06 & 5.4E-06 & 1.3E-06 & 4.9E-06 & 2.3E-06 \\
\iso{42}Ca & 5.3E-06 & 1.1E-06 & 1.8E-05 & 8.3E-06 & 1.5E-05 & 4.7E-06 \\
\iso{43}Ca & 8.4E-05 & 2.3E-05 & 1.1E-04 & 5.3E-05 & 8.0E-05 & 5.3E-05 \\
\iso{44}Ca & 1.4E-06 & 1.7E-07 & 1.3E-06 & 2.3E-07 & 7.9E-07 & 5.2E-07 \\
\iso{45}Sc & 1.6E-05 & 1.8E-06 & 1.5E-05 & 2.1E-06 & 8.2E-06 & 3.8E-06 \\
\iso{44}Ti & 3.3E-02 & 3.2E-03 & 3.1E-02 & 5.8E-03 & 2.0E-02 & 1.3E-02 \\
\iso{46}Ti & 4.0E-06 & 5.8E-07 & 4.2E-06 & 7.8E-04 & 2.8E-06 & 6.9E-05 \\
\iso{47}Ti & 3.7E-04 & 1.8E-05 & 7.4E-04 & 1.8E-03 & 6.0E-04 & 4.6E-04 \\
\iso{48}Ti & 9.4E-05 & 8.0E-07 & 3.5E-04 & 8.8E-05 & 2.7E-04 & 2.2E-04 \\
\iso{49}Ti & 1.4E-07 & 1.5E-09 & 4.6E-07 & 1.3E-07 & 2.6E-07 & 1.2E-07 \\
\iso{47}V & 1.1E-06 & 2.0E-08 & 5.2E-06 & 6.1E-08 & 9.2E-06 & 2.9E-06 \\
\iso{48}V & 3.2E-03 & 2.7E-05 & 1.2E-02 & 3.0E-03 & 9.4E-03 & 7.4E-03 \\
\iso{49}V & 6.0E-05 & 6.2E-07 & 2.0E-04 & 5.5E-05 & 1.1E-04 & 5.2E-05 \\
\iso{51}V & 1.0E-05 & 5.9E-08 & 4.7E-05 & 6.5E-05 & 3.9E-05 & 2.7E-05 \\
\iso{48}Cr & 2.3E-03 & 2.0E-05 & 8.7E-03 & 2.2E-03 & 7.0E-03 & 5.5E-03 \\
\iso{49}Cr & 1.4E-07 & 4.4E-10 & 7.7E-07 & 3.1E-08 & 2.2E-06 & 1.4E-06 \\
\iso{50}Cr & 2.0E-05 & 2.3E-07 & 6.1E-05 & 3.8E-04 & 3.6E-05 & 3.2E-05 \\
\iso{51}Cr & 3.6E-04 & 2.1E-06 & 1.7E-03 & 2.3E-03 & 1.4E-03 & 9.4E-04 \\
\iso{51}Mn & 2.4E-07 & 2.2E-10 & 2.5E-06 & 2.3E-07 & 1.6E-05 & 1.4E-05 \\
\iso{52}Mn & 7.8E-04 & 2.0E-06 & 9.1E-03 & 1.6E-02 & 2.4E-02 & 2.1E-02 \\
\iso{53}Mn & 5.0E-05 & 1.2E-07 & 5.7E-04 & 2.6E-04 & 1.2E-03 & 4.7E-04 \\
\iso{51}Fe & 9.7E-08 & 4.0E-11 & 7.9E-07 & 2.2E-07 & 2.5E-06 & 2.3E-06 \\
\iso{52}Fe & 8.5E-05 & 2.1E-07 & 9.9E-04 & 1.7E-03 & 2.7E-03 & 2.6E-03 \\
\iso{53}Fe & 2.6E-08 & 2.9E-12 & 3.3E-07 & 7.7E-09 & 5.7E-06 & 1.6E-05 \\
\iso{54}Fe & 1.7E-05 & 4.2E-08 & 1.5E-04 & 3.7E-04 & 2.7E-04 & 1.2E-04 \\
\iso{55}Fe & 7.8E-05 & 1.2E-07 & 7.7E-04 & 1.8E-03 & 2.4E-03 & 1.5E-03 \\
\iso{56}Fe & 9.5E-08 & 1.0E-10 & 1.0E-06 & 7.9E-05 & 1.1E-05 & 2.7E-05 \\
\iso{57}Fe & 2.2E-08 & 2.8E-11 & 1.7E-07 & 1.0E-05 & 1.4E-06 & 3.1E-06 \\
\iso{55}Co & 3.9E-05 & 5.8E-08 & 3.8E-04 & 9.1E-04 & 1.2E-03 & 7.8E-04 \\
\iso{56}Co & 1.5E-05 & 1.5E-08 & 1.8E-04 & 1.5E-02 & 2.0E-03 & 5.2E-03 \\
\iso{57}Co & 1.3E-05 & 1.7E-08 & 1.1E-04 & 6.4E-03 & 8.3E-04 & 2.0E-03 \\
\iso{56}Ni & 9.9E-05 & 9.8E-08 & 1.2E-03 & 1.1E-01 & 1.4E-02 & 3.7E-02 \\
\iso{57}Ni & 1.8E-05 & 2.3E-08 & 1.5E-04 & 9.1E-03 & 1.2E-03 & 2.8E-03 \\
\iso{58}Ni & 1.6E-05 & 3.0E-08 & 1.4E-04 & 1.8E-02 & 8.7E-04 & 2.1E-03 \\
\iso{59}Ni & 1.4E-07 & 4.4E-10 & 1.8E-06 & 3.4E-05 & 8.6E-06 & 1.8E-05 \\
\iso{60}Ni & 4.0E-09 & 2.1E-11 & 7.7E-08 & 6.0E-07 & 4.6E-07 & 1.0E-06 \\
\enddata
\enddata
\enddata
\tablecomments{ The products listed include the He layer only, and are given at the beginning of radiative transfer calculation ($t=10^5 s$). Between He and K, only the most abundant isotope is included, from Ca and up all isotopes exceeding $10^{-6} \msun$ in at least one model are listed. EK - final kinetic energy in units of $10^{51}$ ergs, isotopes - in solar mass.}

\end{deluxetable*}

\begin{deluxetable*}{lcl}
\tabletypesize{\scriptsize}
\tablecaption{End product as a function of initial \iso{12}C mass fraction \label{tab:mix}}
\tablehead{
\colhead{$X_C(initial)$}&
\colhead{N}&
\colhead{End-product($\iso{12}C+N \alpha$)}
}
\startdata
       0.6       &      2     &        \iso{20}Ne \\
       0.5       &      3     &        \iso{24}Mg \\
       0.429     &      4     &        \iso{28}Si \\
       0.375     &      5     &        \iso{32}S  \\
       0.33      &      6     &        \iso{36}Ar \\
       0.30      &      7     &        \iso{40}Ca \\
       0.273     &      8     &        \iso{44}Ti \\
       0.25      &      9     &        \iso{48}Cr \\
       0.231     &     10     &        \iso{52}Fe \\
       0.214     &     11     &        \iso{56}Ni \\
\enddata

\end{deluxetable*}

\end{document}